\documentclass[a4paper,aps,prd,10pt,preprintnumbers,showpacs,showkeys,twocolumn,superscriptaddress,nofootinbib,amsmath,amssymb]{revtex4-1}
\usepackage{cmap}
\usepackage[utf8]{inputenc}
\usepackage[T1]{fontenc}
\usepackage{graphicx,orcidlink,xcolor,ulem}

\begin{document}

\title{Gravitational quasinormal modes of Dymnikova black holes}

\author{Bekir Can Lütfüoğlu}
\email{bekir.lutfuoglu@uhk.cz}
\affiliation{Department of Physics, Faculty of Science, University of Hradec Králové, Rokitanského 62/26, 500 03 Hradec Králové, Czech Republic}
\author{Erdinç Ulaş Saka}
\email{ulassaka@istanbul.edu.tr}
\affiliation{Department of Physics, Faculty of Science, Istanbul University, Vezneciler, 34134 Istanbul, Türkiye}

\author{Abubakir Shermatov}
\email{shermatov.abubakir98@gmail.com}
\affiliation{Institute of Fundamental and Applied Research, National Research University TIIAME, Kori Niyoziy 39, Tashkent 100000, Uzbekistan}
\affiliation{University of Tashkent for Applied Sciences, Str. Gavhar 1, Tashkent 100149, Uzbekistan}
\affiliation{Tashkent State Technical University, Tashkent 100095, Uzbekistan}

\author{Javlon Rayimbaev}
\email{javlon@astrin.uz}
\affiliation{
Urgench State University, Kh. Alimjan Str. 14, Urgench 221100, Uzbekistan}


\author{Inomjon Ibragimov}
\email{i.ibragimov@kiut.uz}
\affiliation{Kimyo International University in Tashkent, Shota Rustaveli street 156, Tashkent 100121, Uzbekistan}

\author{Sokhibjan Muminov} \email{sokhibjan.muminov@mamunedu.uz}   \affiliation{Mamun University, Bolkhovuz street 2, Khiva 220900, Uzbekistan } 



\begin{abstract}  
We investigate gravitational quasinormal modes of the Dymnikova black hole, a regular spacetime in which the central singularity is replaced by a de Sitter core. This geometry, originally proposed as a phenomenological model, also arises naturally in the framework of Asymptotically Safe gravity, where quantum corrections lead to a scale-dependent modification of the Schwarzschild solution. Focusing on axial gravitational perturbations, we compute the dominant quasinormal frequencies using the WKB method with Padé approximants and verify the results with time-domain integration. We find that the introduction of the quantum parameter $l_{\rm cr}$ leads to systematic deviations from the Schwarzschild spectrum: the real oscillation frequency decreases as $l_{\rm cr}$ increases, while the damping rate also becomes smaller, implying longer-lived modes. In the limit of large $l_{\rm cr}$, the quasinormal spectrum smoothly approaches the Schwarzschild case. These results suggest that even though the corrections are localized near the horizon, they leave imprints in the gravitational-wave ringdown which may become accessible to observation with future high-precision detectors.  
\end{abstract}  

\maketitle

\section{Introduction}

One of the most powerful probes of black-hole physics comes from the study of their quasinormal modes (QNMs)  \cite{Kokkotas:1999bd,Berti:2009kk,Konoplya:2011qq,Bolokhov:2025uxz}, which govern the damped oscillations of spacetime following a perturbation. These modes encode key information about the underlying geometry and dominate the ringdown phase of gravitational-wave signals detected by LIGO, Virgo, and KAGRA \cite{LIGOScientific:2016aoc, LIGOScientific:2017vwq, LIGOScientific:2020zkf,KAGRA:2013rdx}. The spectrum of gravitational perturbations provides a unique fingerprint of the compact object, enabling tests of general relativity in the strong-field regime and offering the possibility of constraining or ruling out alternative models of gravity and exotic compact objects. For this reason, accurate determination of gravitational QNMs has become a central task in both theoretical and observational astrophysics.  

Alongside these observational advances, increasing attention has been directed toward regular black holes—geometries in which curvature singularities are avoided while preserving the main features of standard black-hole spacetimes. Such models provide a phenomenological window into possible effects of quantum gravity and serve as effective descriptions of singularity resolution. They have been studied from multiple perspectives, including energy conditions, stability, and thermodynamics, and have been confronted with astrophysical data in the context of shadows, gravitational lensing, and quasinormal ringing. In particular, QNM spectra of regular black holes often display deviations from the Schwarzschild case that can, in principle, leave observable imprints in gravitational-wave signals.  

Quantum effects are likely to leave their imprint not only on the geometry but also on the behavior of test particles, potentially altering aspects of their dynamics \cite{Turimov:2025tmf}. Moreover, in scenarios where matter fields interact with the background in a non-minimal fashion, or when analyzing associated radiation phenomena, the role of boundary conditions may also acquire particular meaning \cite{2025JPhA...58H5201D,2025EPJB...98...35R}. 

Among the various proposals, the solution introduced by Dymnikova \cite{Dymnikova:1992ux} stands out as an elegant example. It describes a static, spherically symmetric, asymptotically flat geometry in which the central singularity is replaced by a de Sitter core, leading to finite curvature invariants everywhere. While originally formulated as a phenomenological construction, the same functional form has more recently reappeared in the framework of Asymptotic Safety, where it can be derived from renormalization-group improvements tied to curvature invariants \cite{Platania:2019kyx}. Thus, the Dymnikova spacetime provides not only one of the earliest regular black-hole models, but also a concrete realization of quantum-gravity inspired corrections. This dual interpretation makes it a natural and timely candidate for detailed studies of gravitational perturbations and their observational consequences.  

QNMs of test fields in Dymnikova black holes have been investigated in four and higher dimensions in several recent works~\cite{Konoplya:2024kih,Macedo:2024dqb,Konoplya:2023aph}. However, to the best of our knowledge, no analogous analysis has yet been carried out for the most important case of gravitational perturbations within the framework of Asymptotically Safe gravity. In this paper we address this gap by computing the dominant QNMs of gravitational perturbations in the Dymnikova background. It is worth noting that grey-body factors and absorption cross-sections for gravitational perturbations of the Dymnikova solution were recently computed in~\cite{Dubinsky:2025nxv}. However, that analysis did not provide any results for the quasinormal frequencies.  

It is worth mentioning that the literature on QNMs of quantum-corrected black holes is vast; therefore, we cite here only a few  works~\cite{Konoplya:2020der,Chen:2023wkq,Baruah:2023rhd,Moreira:2023cxy,Fu:2023drp,Heidari:2023ssx,Skvortsova:2024msa,Kokkotas:2017zwt,Abdalla:2005hu,Skvortsova:2023zca,Bonanno:2025dry,Bolokhov:2023ozp,Bolokhov:2025lnt,Malik:2024elk,Malik:2025dxn}, referring the reader to the references therein for further studies.

The paper is organized as follows. In Sec.~\ref{sec:dymnikova} we introduce the Dymnikova black-hole geometry in the context of Asymptotic Safety. Sec.~\ref{sec:perturbations} is devoted to the analysis of axial gravitational perturbations and the derivation of the effective potential. In Sec.~\ref{QNmodes} we describe the numerical techniques used for calculating QNMs, including the WKB method with Padé approximants and the time-domain integration scheme. The obtained results for the quasinormal spectrum are presented and discussed in Sec.~\ref{QNmodes}. Finally, in Sec.~\ref{Concl} we summarize our findings and outline possible directions for future research.

\section{The Dymnikova Black Hole Geometry}\label{sec:dymnikova}

The quest for singularity-free black-hole solutions has long motivated the construction of effective geometries that remain regular throughout spacetime. A particularly influential example was proposed by Dymnikova~\cite{Dymnikova:1992ux}, who suggested replacing the central singularity of the Schwarzschild solution with a de Sitter core while preserving asymptotic flatness. This is achieved through a stress--energy distribution that formally violates some of the classical energy conditions but can be regarded as an effective description of quantum backreaction. The outcome is a static, spherically symmetric, and asymptotically flat spacetime in which all curvature invariants remain finite.  

The line element retains the usual Schwarzschild form,
\begin{equation}\label{metric}
ds^2 = -f(r)\, dt^2 + \frac{dr^2}{f(r)} + r^2 \bigl(d\theta^2 + \sin^2\theta\, d\phi^2\bigr),
\end{equation}
but with a modified lapse function
\begin{equation}
f(r) = 1 - \frac{2 M(r)}{r},
\end{equation}
where the mass function interpolates between zero at the origin and the asymptotic ADM mass $M$ at infinity. Dymnikova’s original proposal was
\begin{equation}
M(r) = M\Bigl(1 - e^{-r^3/r_0^3}\Bigr),
\end{equation}
with $r_0$ characterizing the size of the de Sitter core. At small $r$, the geometry tends to
\[
f(r) \approx 1 - \frac{r^2}{r_0^2},
\]
describing a regular de Sitter interior, while at large $r$ one recovers the Schwarzschild behavior $f(r) \approx 1 - 2M/r$. The spacetime thus interpolates smoothly between two familiar limits while remaining free of singularities.

Although originally introduced phenomenologically, essentially the same functional form was later derived within the framework of Asymptotic Safety. In this approach, the Schwarzschild lapse function
\begin{equation}\label{lapse}
f(r) = 1 - \frac{2 M}{r},
\end{equation}
is modified by allowing Newton’s constant to run with the radial scale. A convenient parametrization is
\begin{equation}\label{runningG}
G(r) = \frac{G_0}{1 + g_\ast^{-1} G_0 k^2(r)},
\end{equation}
where $G_0$ is the classical Newton coupling, $g_\ast$ denotes the fixed-point value, and $k(r)$ is an effective cutoff scale satisfying $k(r)\!\to\!0$ as $r\!\to\!\infty$, thus ensuring the recovery of classical Schwarzschild behavior at large distances. Substituting $G(r)$ into the metric leads to
\begin{equation}
f(r) = 1 - \frac{2 M}{r}\,\frac{G(r)}{G_0}.
\end{equation}

The modified lapse can be reinterpreted as arising from an effective energy--momentum tensor,
\begin{equation}\label{Teff}
T_{\mu\nu} = p g_{\mu\nu} + (\rho+p)(l_\mu n_\nu + l_\nu n_\mu) ,
\end{equation}
where $l_\mu n^\mu = -1$. The effective energy density and pressure are generated by the radial dependence of $G(r)$:
\begin{equation}\label{rho-p}
p = -\frac{M G''(r)}{8 \pi r\, G(r)},
\qquad 
\rho = \frac{M G'(r)}{4 \pi r^2 G(r)}.
\end{equation}
This stress tensor can be interpreted as the imprint of quantum vacuum polarization: deviations of $G(r)$ from constancy induce effective matter sources that regularize the geometry.

A self-consistent solution is obtained through an iterative renormalization-group improvement: starting with the classical Schwarzschild case ($G(r)=G_0$), one updates the cutoff $k(r)$ as a functional of the energy density generated at the previous step. In the continuum limit, this iterative process converges to a closed analytic expression for the lapse, which coincides with the Dymnikova form~\cite{Platania:2019kyx}:
\begin{equation}\label{dymnikova-final}
f(r) = 1 - \frac{2 M}{r}\left(1 - e^{-r^3/(2 l_{\rm cr}^2 M)}\right).
\end{equation}
Here $l_{\rm cr}$ is a critical length scale marking the onset of quantum corrections. For $l_{\rm cr}\!\to\!0$ the Schwarzschild metric is recovered, while finite $l_{\rm cr}$ introduces a de Sitter-like core. The existence of an event horizon requires
\[
l_{\rm cr} \lesssim 1.138\,M,
\]
with larger values of $l_{\rm cr}$ corresponding instead to horizonless, compact, but nonsingular configurations.  From now on, we will use units $M=1$. Thus, $\omega$ implies $\omega M$ in dimensionless units.

The Dymnikova solution thus enjoys a dual origin: historically as a phenomenological model of singularity resolution and, more recently, as a concrete realization of renormalization-group improvements in Asymptotic Safety. This dual perspective underscores its relevance as a minimal yet robust model of a regular black hole. Furthermore, higher-dimensional generalizations of the Dymnikova geometry have been shown to arise naturally in theories with higher-curvature corrections \cite{Konoplya:2024kih}, further highlighting its role as a testing ground for semiclassical and quantum-gravity inspired modifications of black-hole physics.  

\section{Gravitational perturbations}\label{sec:perturbations}

\begin{figure}
\resizebox{\linewidth}{!}{\includegraphics{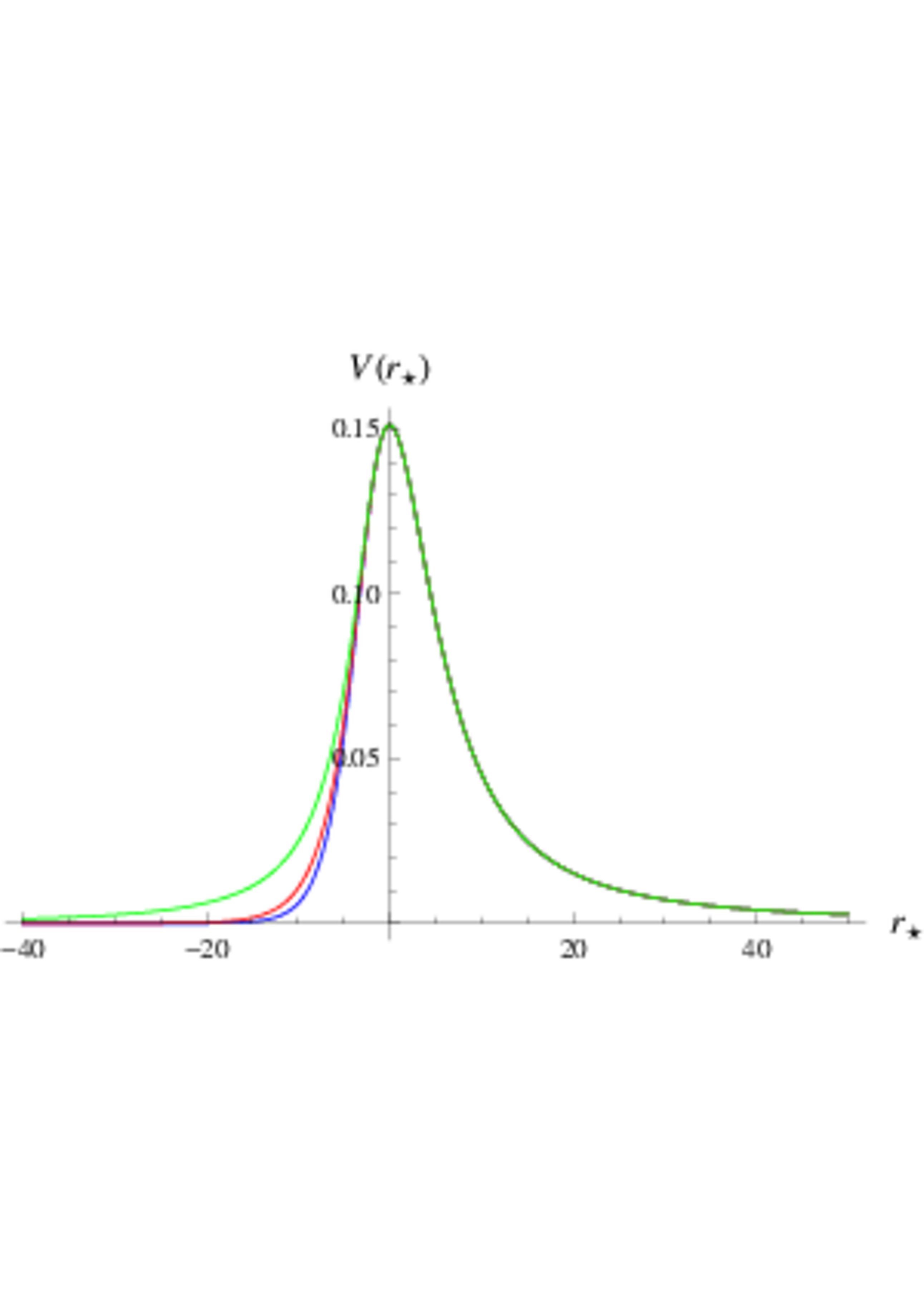}}
\caption{Effective potential as a function of the tortoise coordinate $r^{*}$ for $\ell=2$, $l_{cr}=0.1$ (blue), $l_{cr}=1$ (red), $l_{cr}=1.137$ (green).}\label{fig:gravpot1}
\end{figure}

\begin{figure}
\resizebox{\linewidth}{!}{\includegraphics{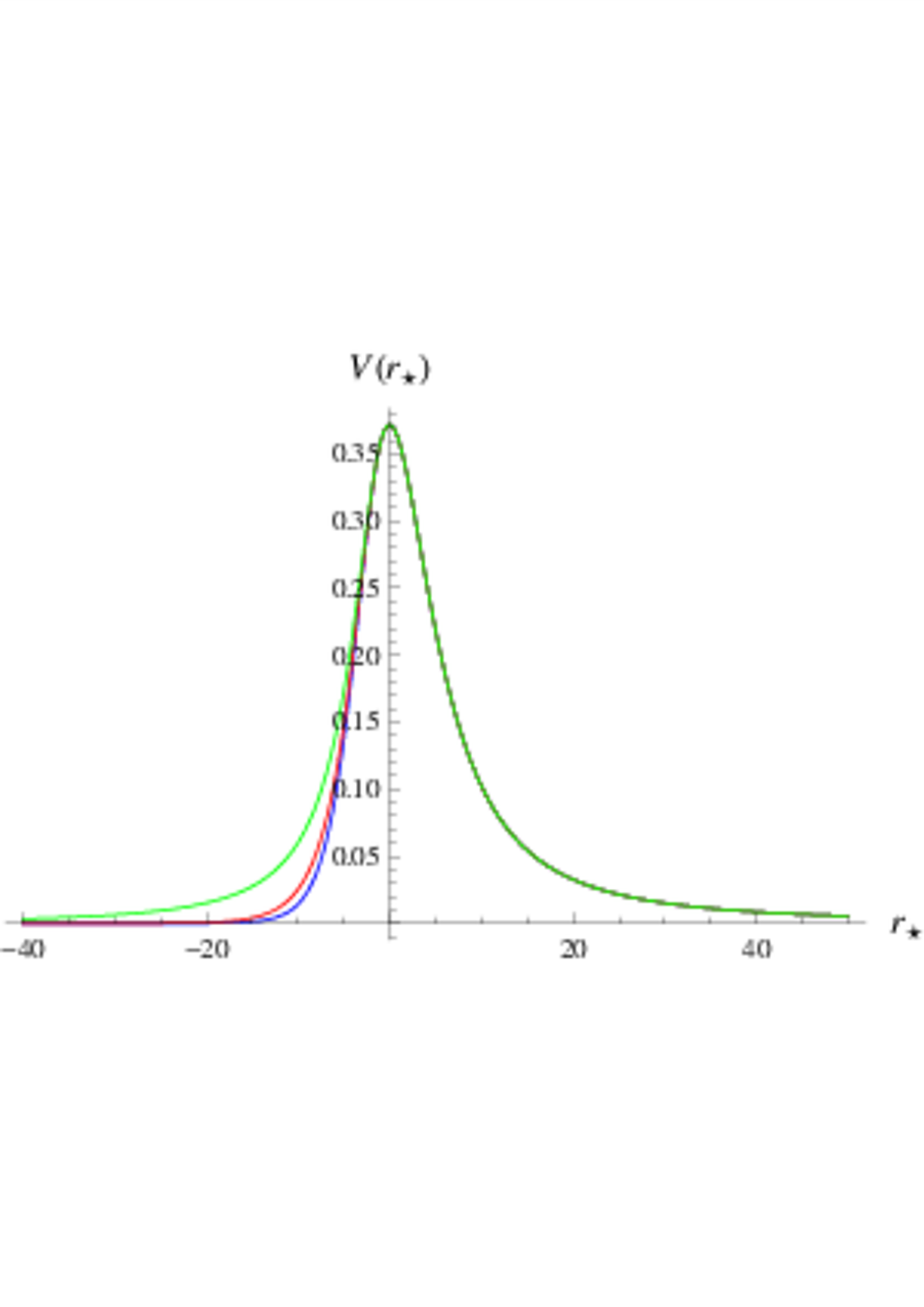}}
\caption{Effective potential as a function of the tortoise coordinate $r^{*}$ for $\ell=3$, $l_{cr}=0.1$ (blue), $l_{cr}=1$ (red), $l_{cr}=1.137$  (green).}\label{fig:gravpot2}
\end{figure}

Studying gravitational perturbations in spacetimes that incorporate quantum corrections is more subtle than in the classical case. The difficulty arises because such geometries, including the Dymnikova black hole, are not obtained as exact solutions of Einstein’s equations with a specified matter source, but rather emerge from effective constructions motivated by quantum-gravity considerations. As a result, a straightforward derivation of perturbation equations is not possible. A practical resolution was suggested in the loop-inspired framework of Ashtekar, Olmedo, and Singh~\cite{Ashtekar:2018lag,Ashtekar:2018cay}, where the background geometry can be reinterpreted within Einstein’s theory if one introduces an effective anisotropic fluid stress tensor that captures the relevant quantum corrections. This reformulation allows the machinery of black-hole perturbation theory to be applied in a consistent manner.  

Following this idea, Bouhmadi-López and collaborators~\cite{Bouhmadi-Lopez:2020oia,Konoplya:2024lch} analyzed axial perturbations under the assumption that fluctuations aligned with the anisotropy do not contribute in the axial sector. This simplification is reminiscent of the treatment in~\cite{Bronnikov:2012ch}, where the decoupling of scalar and axial gravitational modes was exploited in scalar–tensor models (see also~\cite{Chen:2019iuo}). Although such assumptions necessarily restrict the analysis, they have been widely adopted in related contexts and are generally expected to provide reliable results for small deviations from Schwarzschild geometry \cite{Konoplya:2025hgp,Konoplya:2024lch,Dubinsky:2025nxv,Bolokhov:2025egl}. {For Dymnikova black holes as solutions of the Einstein equations with the energy-momentum tensor corresponding to some density profile, the perturbation equations were considered in \cite{Dymnikova:2004qg}.}

Under these conditions, the perturbation equations can be reduced to a master wave equation of Schrödinger type,
\begin{equation}
\frac{d^2\Psi}{dr_{*}^2} + \bigl(\omega^2 - V(r)\bigr)\Psi = 0,
\end{equation}
with the tortoise coordinate defined by $dr^*/dr = 1/f(r)$. The effective axial potential takes the form
\begin{equation}
V(r) = f(r)\left[\frac{(\ell+2)(\ell-1)}{r^2} - \frac{f'(r)}{r} + \frac{2 f(r)}{r^2}\right],
\end{equation}
where $\ell=2,3,4,\ldots$ labels the multipole number. Unlike the standard Regge–Wheeler potential, this expression incorporates contributions from perturbations of the effective fluid stress tensor. Consequently, it deviates from the vacuum Schwarzschild case even when $f(r)$ reduces to the Schwarzschild form.

Figures~\ref{fig:gravpot1} and \ref{fig:gravpot2} show the effective potentials for axial modes in the Dymnikova background. When the quantum parameter $l_{\rm cr}$ is very small (e.g., $l_{\rm cr}=0.01M$), the potential nearly coincides with the Schwarzschild one except in the immediate vicinity of the horizon. The deformation is strongly localized, and the potential rapidly merges with the Schwarzschild profile at moderate radii. At the same time, the location of the event horizon shifts inward as $l_{\rm cr}$ grows, leading to a decrease in the horizon radius compared to the Schwarzschild value. These features indicate that the quantum-inspired modifications are concentrated in the near-horizon region, leaving the asymptotic structure essentially unaffected.

Analogous simplifications have been applied in several earlier studies of black-hole perturbations~\cite{Berti:2003yr,Kokkotas:1993ef,Konoplya:2006ar}. While such approaches neglect certain couplings that might alter the detailed spectrum, they capture the leading behavior whenever the corrections relative to Schwarzschild are small. This perspective fits well with the perturbative character of quantum-gravity modifications, which are expected to act only as moderate deformations of the classical background.

\section{Quasinormal modes}\label{QNmodes}

For black holes, $V(r)$ has a single peak at $r=r_0$ (or $r_*=r_{*0}$), vanishes exponentially as $r_*\to-\infty$ (event horizon), and falls off as $r_*\to+\infty$ (spatial infinity).  QNMs are defined by the purely ingoing/outgoing asymptotics
\begin{eqnarray}
\Psi \propto e^{-i\omega r_*},\quad r_*\to -\infty \quad (\text{horizon}),
\\
\Psi \propto e^{+i\omega r_*},\quad r_*\to +\infty \quad (\text{infinity})\,.
\label{eq:BCs}
\end{eqnarray}
These conditions select a discrete set of complex frequencies $\omega=\omega_R-i\omega_I$ with $\omega_I>0$. Following the Schutz–Will/Iyer–Will approach~\cite{Schutz:1985km,Iyer:1986np} refined to higher orders~\cite{Konoplya:2003ii,Matyjasek:2017psv,Konoplya:2019hlu}, define
\begin{equation}
Q(r_*)\equiv \omega^{2}-V(r_*), \quad
V_0 \equiv V(r_{*0}),\quad V_0^{(k)}\equiv \left.\frac{d^{k}V}{dr_*^{k}}\right|_{r_{*0}}.
\end{equation}
Expanding around the peak and matching WKB solutions across the turning points yields the $N$-th order quantization condition
\begin{equation}
\frac{i\bigl(\omega^{2}-V_0\bigr)}{\sqrt{-2V_0^{(2))}}}
-\sum_{k=2}^{N}\Lambda_{k}\!\left(\{V_0^{(j)}\},n\right)
= n+\frac{1}{2}\,, n=0,1,2,\ldots,
\label{eq:WKB-N}
\end{equation}
where primes denote $r_*$-derivatives, and the $\Lambda_k$ are explicit polynomials in $V_0^{(j)}$ ($2<j\le 2k$) and $n$, divided by the corresponding powers of $V_0^{(2)}$ (see \cite{Iyer:1986np,Konoplya:2003ii,Matyjasek:2017psv,Konoplya:2019hlu} for closed forms up to high orders). 

Over the past decades, the WKB approach and its refinements have become one of the most widely employed semi-analytical tools for studying black-hole spectra. Its applications extend far beyond the Schwarzschild case, encompassing higher-curvature theories, quantum-corrected geometries, and a variety of exotic compact objects. Numerous examples of such developments can be found in~\cite{Matyjasek:2021xfg,Pedrotti:2025idg,Konoplya:2020cbv,MahdavianYekta:2019pol,Zhang:2025xqt,Antonelli:2025yol,Lutfuoglu:2025ljm,Hamil:2025cms,Miyachi:2025ptm,Matyjasek:2017psv,Konoplya:2005sy,Konoplya:2021ube,Konoplya:2023moy}, which collectively demonstrate both the flexibility and robustness of the method across different gravitational settings.  

In the present analysis, our focus is on extracting the two lowest-lying quasinormal frequencies, $\omega_{0}$ and $\omega_{1}$. For this purpose, we employ the sixth or higher-order WKB formalism supplemented by Padé approximants with indices $\tilde m = \tilde n = 3$~\cite{Konoplya:2019hlu}. This refinement is known to provide optimal accuracy in many situations, as supported by a wide range of recent investigations~\cite{Dubinsky:2025fwv,Skvortsova:2023zmj,Bronnikov:2019sbx,Konoplya:2019xmn,Bolokhov:2024ixe,Dubinsky:2025azv,Skvortsova:2024atk,Dubinsky:2024hmn,Malik:2023bxc,Churilova:2021tgn,Lutfuoglu:2025hjy,Dubinsky:2024rvf,Malik:2024nhy,Dubinsky:2025bvf,Skvortsova:2024wly,Bolokhov:2023bwm,Dubinsky:2024nzo}. The convergence of results across these diverse studies highlights the reliability of the WKB–Padé scheme for accurately capturing the dominant part of the quasinormal spectrum.  

It is convenient to introduce a bookkeeping parameter $\epsilon$ (set to 1 at the end) and write
\begin{equation}
\frac{i\bigl(\omega^{2}-V_0\bigr)}{\epsilon\sqrt{-2V_0^{(2)}}}
-\sum_{k=2}^{N}\epsilon^{k-1}\Lambda_{k}= n+\frac{1}{2}\,,
\label{eq:WKB-epsilon}
\end{equation}
which generates a formal WKB series for $\omega^{2}$,
\begin{equation}
\omega^{2}(\epsilon)=V_0 - i\epsilon\Bigl(n+\tfrac{1}{2}\Bigr)\sqrt{-2V_0^{(2)}}
+\sum_{j=2}^{N} c_{j}\,\epsilon^j\,,
\label{eq:omega-series}
\end{equation}
with the coefficients $c_j=c_j\bigl(n,V_0^{(2)},V_0^{(3)},\ldots,V_0^{(2j)}\bigr)$ obtained algebraically from~\eqref{eq:WKB-epsilon}.

Because the WKB series is asymptotic, accuracy improves markedly by applying a Padé resummation to the truncated polynomial $P_N(\epsilon)\equiv \omega^{2}(\epsilon)$:
\begin{equation}
\mathcal{P}_{\tilde m/\tilde n}(\epsilon)
=\frac{\displaystyle \sum_{j=0}^{\tilde m} a_j\,\epsilon^{j}}
{\displaystyle 1+\sum_{k=1}^{\tilde n} b_k\,\epsilon^{k}}
\qquad (\tilde m+\tilde n=N)\,,
\label{eq:pade}
\end{equation}
with $a_j,b_k$ fixed by matching the Taylor expansion of $\mathcal{P}_{\tilde m/\tilde n}$ to $P_N$ up to $\mathcal{O}(\epsilon^{N+1})$. The Padé estimate for the QNM is then
\begin{equation}
\omega^2 = \mathcal{P}_{\tilde m/\tilde n}(1)\,,
\qquad
\omega=\sqrt{\mathcal{P}_{\tilde m/\tilde n}(1)}\,,
\label{eq:omega-pade}
\end{equation}
where the principal square root is chosen by continuity from the eikonal limit. In practice we use balanced or near-balanced approximants, e.g.\ $[\tilde m/\tilde n]=[3/3],[4/4]$ for $N=7,9$, etc., and quote the spread among a small family of $(\tilde m,\tilde n)$ as an internal error estimate.

We use the following steps to find QNMs:
\begin{enumerate}
\item Locate the potential maximum $r_0$ by solving $V'(r_0)=0$; compute $r_{*0}$ and $V_0^{(k)}$ up to the order needed by $N$.
\item Build the WKB quantization condition~\eqref{eq:WKB-N} and extract the truncated series $P_N(\epsilon)$ for $\omega^{2}$ as in~\eqref{eq:omega-series}.
\item Construct Padé approximants $\mathcal{P}_{\tilde m/\tilde n}(\epsilon)$ per~\eqref{eq:pade} and evaluate at $\epsilon=1$ to obtain $\omega$ via~\eqref{eq:omega-pade}.
\item Validate by varying $N$ and $(\tilde m,\tilde n)$, and by checking stability against small changes of the numerical derivatives $V_0^{(k)}$.
\end{enumerate}

The WKB–Padé method is most reliable for moderate to large multipoles and low overtones (heuristically $n\ll \ell$ for single-barrier potentials). Tables I–{VIII} summarize the results obtained with 6th and higher-order WKB methods combined with Padé resummation, while selected modes are validated through time-domain extraction when necessary.

\begin{table}[h!]
\begin{tabular}{c c c c}
\hline
\hline
$l_{cr}$ & WKB-6 $m=3$ & WKB-7 $m=4$ & difference  \\
\hline
$0.3$ & $0.373642-0.088936 i$ & $0.373642-0.088957 i$ & $0.00536\%$\\
$0.5$ & $0.373642-0.088936 i$ & $0.373659-0.088971 i$ & $0.0102\%$\\
$0.7$ & $0.373446-0.089169 i$ & $0.373723-0.088886 i$ & $0.103\%$\\
$0.9$ & $0.374020-0.089779 i$ & $0.373285-0.088740 i$ & $0.331\%$\\
$1.13$ & $0.369187-0.085818 i$ & $0.369142-0.085698 i$ & $0.0338\%$\\
$1.137$ & $0.368999-0.085650 i$ & $0.368956-0.085529 i$ & $0.0341\%$\\
\hline
\hline
\end{tabular}
\caption{QNMs of the $\ell=2$, $n=0$ test axial gravitational perturbations for the Dymnikova black hole $M=1$ calculated using the WKB formula at different orders and Padé approximants.}
\end{table}

\begin{table}[h!]
\begin{tabular}{c c c c}
\hline
\hline
$l_{cr}$ & WKB-6 $m=3$ & WKB-7 $m=4$ & difference  \\
\hline
$0.3$ & $0.346099-0.273533 i$ & $0.345935-0.274690 i$ & $0.265\%$\\
$0.5$ & $0.346099-0.273533 i$ & $0.346769-0.274310 i$ & $0.233\%$\\
$0.7$ & $0.348365-0.277361 i$ & $0.345919-0.272993 i$ & $1.12\%$\\
$0.9$ & $0.350747-0.275420 i$ & $0.344435-0.270005 i$ & $1.86\%$\\
$1.13$ & $0.325487-0.257860 i$ & $0.324701-0.256776 i$ & $0.322\%$\\
$1.137$ & $0.324627-0.257215 i$ & $0.323942-0.256233 i$ & $0.289\%$\\
\hline
\hline
\end{tabular}
\caption{QNMs of the $\ell=2$, $n=1$ for the Dymnikova black hole $M=1$ calculated using the WKB formula at different orders and Padé approximants.}
\end{table}

\begin{table}[h!]
\begin{tabular}{c c c c}
\hline
\hline
$l_{cr}$ & WKB-6 $m=3$ & WKB-7 $m=4$ & difference  \\
\hline
$0.3$ & $0.599443-0.092703 i$ & $0.599443-0.092703 i$ & $0\%$\\
$0.5$ & $0.599443-0.092703 i$ & $0.599443-0.092703 i$ & $0\%$\\
$0.7$ & $0.599465-0.092675 i$ & $0.599454-0.092707 i$ & $0.00547\%$\\
$0.9$ & $0.598895-0.092830 i$ & $0.598987-0.092770 i$ & $0.0181\%$\\
$1.13$ & $0.596184-0.091107 i$ & $0.596288-0.091159 i$ & $0.0193\%$\\
$1.137$ & $0.595927-0.090979 i$ & $0.596149-0.091043 i$ & $0.0384\%$\\
\hline
\hline
\end{tabular}
\caption{QNMs of the $\ell=3$, $n=0$ for the Dymnikova black hole $M=1$ calculated using the WKB formula at different orders and Padé approximants.}
\end{table}

\begin{table}[h!]
\begin{tabular}{c c c c}
\hline
\hline
$l_{cr}$ & WKB-6 $m=3$ & WKB-7 $m=4$ & difference  \\
\hline
$0.3$ & $0.582642-0.281298 i$ & $0.582645-0.281299 i$ & $0.00037\%$\\
$0.5$ & $0.582642-0.281298 i$ & $0.582644-0.281297 i$ & $0.00026\%$\\
$0.7$ & $0.582612-0.280989 i$ & $0.582775-0.281350 i$ & $0.0613\%$\\
$0.9$ & $0.578793-0.281739 i$ & $0.579723-0.281044 i$ & $0.180\%$\\
$1.13$ & $0.568787-0.275578 i$ & $0.567908-0.274439 i$ & $0.228\%$\\
$1.137$ & $0.567875-0.274646 i$ & $0.567306-0.274071 i$ & $0.128\%$\\
\hline
\hline
\end{tabular}
\caption{QNMs of the $\ell=3$, $n=1$  for the Dymnikova black hole $M=1$ calculated using the WKB formula at different orders and Padé approximants.}
\end{table}

\begin{table}[h!]
\begin{tabular}{c c c c}
\hline
\hline
$l_{cr}$ & WKB-6 $m=3$ & WKB-7 $m=4$ & difference  \\
\hline
$0.3$ & $0.809178-0.094164 i$ & $0.809178-0.094164 i$ & $0\%$\\
$0.5$ & $0.809178-0.094164 i$ & $0.809178-0.094164 i$ & $0\%$\\
$0.7$ & $0.809185-0.094154 i$ & $0.809187-0.094169 i$ & $0.00193\%$\\
$0.9$ & $0.808894-0.094354 i$ & $0.808915-0.094290 i$ & $0.00821\%$\\
$1.13$ & $0.806684-0.093368 i$ & $0.806746-0.093185 i$ & $0.0239\%$\\
$1.137$ & $0.806709-0.093176 i$ & $0.806634-0.093089 i$ & $0.0142\%$\\
\hline
\hline
\end{tabular}
\caption{QNMs of the $\ell=4$, $n=0$  for the Dymnikova black hole $M=1$ calculated using the WKB formula at different orders and Padé approximants.}
\end{table}

\begin{table}[h!]
\begin{tabular}{c c c c}
\hline
\hline
$l_{cr}$ & WKB-6 $m=3$ & WKB-7 $m=4$ & difference  \\
\hline
$0.3$ & $0.796631-0.284334 i$ & $0.796632-0.284334 i$ & $0.00003\%$\\
$0.5$ & $0.796631-0.284334 i$ & $0.796631-0.284334 i$ & $0.00002\%$\\
$0.7$ & $0.796637-0.284213 i$ & $0.796725-0.284362 i$ & $0.0204\%$\\
$0.9$ & $0.794730-0.285220 i$ & $0.794916-0.284828 i$ & $0.0515\%$\\
$1.13$ & $0.785538-0.280464 i$ & $0.785165-0.280327 i$ & $0.0476\%$\\
$1.137$ & $0.784954-0.280114 i$ & $0.784672-0.280018 i$ & $0.0357\%$\\
\hline
\hline
\end{tabular}
\caption{QNMs of the $\ell=4$, $n=1$  for the Dymnikova black hole $M=1$ calculated using the WKB formula at different orders and Padé approximants.}
\end{table}

\begin{table}[h!]
\begin{tabular}{c c c c}
\hline
\hline
$l_{cr}$ & WKB-6 $m=3$ & WKB-7 $m=4$ & difference  \\
\hline
$0.3$ & $0.772710-0.479903 i$ & $0.772712-0.479908 i$ & $0.00060\%$\\
$0.5$ & $0.772710-0.479903 i$ & $0.772708-0.479906 i$ & $0.00039\%$\\
$0.7$ & $0.772414-0.479325 i$ & $0.773160-0.479874 i$ & $0.102\%$\\
$0.9$ & $0.764889-0.481360 i$ & $0.765338-0.480293 i$ & $0.128\%$\\
$1.13$ & $0.744544-0.468344 i$ & $0.741081-0.468153 i$ & $0.394\%$\\
$1.137$ & $0.743163-0.467698 i$ & $0.739920-0.467587 i$ & $0.370\%$\\
\hline
\hline
\end{tabular}
\caption{QNMs of the $\ell=4$, $n=2$  for the Dymnikova black hole $M=1$ calculated using the WKB formula at different orders and Padé approximants.}
\end{table}

\begin{table}[h!]
\begin{tabular}{c c c c}
\hline
\hline
$\ell$ & WKB-6 $m=3$ & WKB-7 $m=4$ & difference  \\
\hline
$2$ & $0.368999-0.085650 i$ & $0.368956-0.085529 i$ & $0.0341\%$\\
$3$ & $0.595927-0.090979 i$ & $0.596149-0.091043 i$ & $0.0384\%$\\
$4$ & $0.806709-0.093176 i$ & $0.806634-0.093089 i$ & $0.0142\%$\\
$5$ & $1.010292-0.094095 i$ & $1.010246-0.094067 i$ & $0.00535\%$\\
$6$ & $1.210341-0.094617 i$ & $1.210310-0.094606 i$ & $0.00266\%$\\
$7$ & $1.408314-0.094934 i$ & $1.408295-0.094930 i$ & $0.00143\%$\\
$8$ & $1.604965-0.095140 i$ & $1.604952-0.095139 i$ & $0.00080\%$\\
$9$ & $1.800718-0.095280 i$ & $1.800709-0.095281 i$ & $0.00047\%$\\
$10$ & $1.995834-0.095379 i$ & $1.995829-0.095380 i$ & $0.00028\%$\\
$15$ & $2.966222-0.095605 i$ & $2.966221-0.095606 i$ & $0.00003\%$\\
$20$ & $3.932629-0.095679 i$ & $3.932629-0.095679 i$ & $5.4\times 10^{-6}\%$\\
$25$ & $4.897409-0.095711 i$ & $4.897409-0.095711 i$ & $1.2\times 10^{-6}\%$\\
$30$ & $5.861366-0.095728 i$ & $5.861366-0.095728 i$ & $0\%$\\
$35$ & $6.824849-0.095738 i$ & $6.824849-0.095738 i$ & $0\%$\\
$40$ & $7.788034-0.095744 i$ & $7.788034-0.095744 i$ & $0\%$\\
$45$ & $8.751020-0.095748 i$ & $8.751020-0.095748 i$ & $0\%$\\
$50$ & $9.713865-0.095751 i$ & $9.713865-0.095751 i$ & $0\%$\\
$75$ & $14.526964-0.095758 i$ & $14.526964-0.095758 i$ & $0\%$\\
$100$ & $19.339214-0.095761 i$ & $19.339214-0.095761 i$ & $0\%$\\
$150$ & $28.962861-0.095762 i$ & $28.962861-0.095762 i$ & $0\%$\\
$200$ & $38.586078-0.095763 i$ & $38.586078-0.095763 i$ & $0\%$\\
\hline
\hline
\end{tabular}
\caption{Quasinormal modes for various $\ell$, calculated using the WKB formula at different orders and Pade approximants. Here we have $n=0$, $M=1$, $l_{cr}=1.137$.}\label{Tab8}
\end{table}

An alternative approach to computing QNMs is based on the direct integration of the perturbation equation in the time domain. The wave-like equation containing the second derivative in time instead of $-\omega^2$ is discretized on a characteristic grid using the light-cone variables $u=t-r_*$ and $v=t+r_*$, following the scheme introduced in~\cite{Gundlach:1993tp}. This method evolves an initial pulse through the effective potential and yields the complete time-dependent signal at a fixed spatial location outside the black hole. At intermediate times, the signal is dominated by exponentially damped oscillations corresponding to the quasinormal ringing.  

To extract the dominant frequencies from the numerically obtained waveform, one typically fits the signal to a superposition of damped exponentials. In practice, this is efficiently accomplished with the Prony method, which provides both the real oscillation frequencies and the damping rates. The time-domain approach thus serves as an important cross-check of the WKB–Padé results, and it is particularly useful in regimes where the WKB expansion becomes less reliable.  The time-domain integration together with the Prony method has been described in numerous publications (see, for instance, \cite{Ishihara:2008re,Malik:2025ava,Lutfuoglu:2025bsf,Churilova:2019qph,Konoplya:2013sba,Qian:2022kaq,Varghese:2011ku,Malik:2024bmp,Konoplya:2020jgt,Dubinsky:2024jqi,Cuyubamba:2016cug,Bolokhov:2023ruj,Lutfuoglu:2025hwh,Lutfuoglu:2025qkt}
 for details).

More explicitly, in the light-cone coordinates $u=t-r_*$ and $v=t+r_*$ the wave equation takes the form
\begin{equation}
\frac{\partial^2 \Psi}{\partial u\,\partial v} + \frac{1}{4}V(r)\Psi = 0 .
\end{equation}
Discretizing this equation on a uniform $(u,v)$ grid with step size $\Delta$, one obtains the standard second-order accurate finite-difference scheme \cite{Gundlach:1993tp}
\begin{equation}
\Psi_{N} = \Psi_{W} + \Psi_{E} - \Psi_{S}
- \frac{\Delta^2}{8}\,V\!\left(r_{*}\right)\left(\Psi_{W} + \Psi_{E}\right),
\end{equation}
where the subscripts denote the grid points $S=(u,v)$, $E=(u,v+\Delta)$, $W=(u+\Delta,v)$, and $N=(u+\Delta,v+\Delta)$. The effective potential $V(r)$ is evaluated at the center of the grid cell, corresponding to the tortoise coordinate $r_*(u+\Delta/2,v+\Delta/2)$. This scheme is conditionally stable and converges quadratically with respect to the grid spacing.

The initial data are specified by prescribing a compact Gaussian pulse on one of the null surfaces, for example,
\begin{equation}
\Psi(u=u_0,v) = \exp\!\left[-\frac{(v-v_c)^2}{\sigma^2}\right],
\end{equation}
while setting $\Psi(v=v_0,u)=0$ on the other surface. The subsequent evolution uniquely determines the waveform $\Psi(t,r_*)$ at any fixed observation point outside the black hole. At intermediate times the signal is dominated by exponentially damped oscillations corresponding to the quasinormal ringing, while at late times it transitions to power-law  tails.

To extract the dominant quasinormal frequencies, the time-domain signal in the ringdown regime is fitted to a superposition of damped exponentials,
\begin{equation}
\Psi(t) \simeq \sum_{k=1}^{N} C_k e^{-i\omega_k t},
\end{equation}
using the Prony method. This technique reconstructs the complex frequencies $\omega_k$ by solving a linear prediction problem for a discrete time series sampled at equal time intervals. In practice, the fitting window is chosen such that the prompt response has decayed and the late-time tail has not yet become dominant. The stability of the extracted frequencies is verified by varying the fitting interval and the number of included modes. While the overtones are increasingly difficult to resolve in the time domain, the Prony method reliably captures the fundamental mode.

When using high accuracy time-domain integration, for $\ell=2$, the time-domain integration yields 
\[
\omega = 0.369372 - 0.0856535 i
\]
for the least damped mode. This value is extracted by the Prony method with all accuracy of all 6 digits when fitting at the intervals $t=(150, 200)$ and $t=(200, 250)$. Thus, there is no point in extending computations to later times. The sixth-order WKB method with Padé approximants gives 
\[
\omega = 0.368999 - 0.085650 i.
\]

\begin{figure}[h!]
\resizebox{\linewidth}{!}{\includegraphics{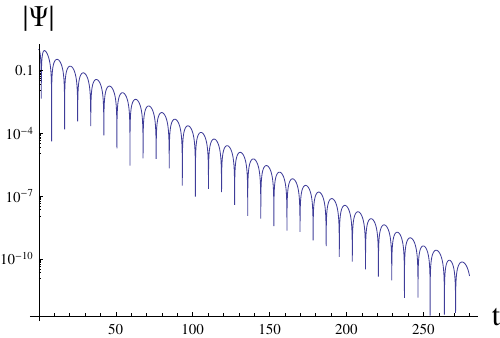}}
\caption{Time-domain profile  for $\ell=2$, $l_{cr}=1.137$. The Prony method gives $\omega = 0.369372 - 0.0856535 i$ for the fundamental mode $n=0$.}\label{fig:L2}
\end{figure}

\begin{figure}
\resizebox{\linewidth}{!}{\includegraphics{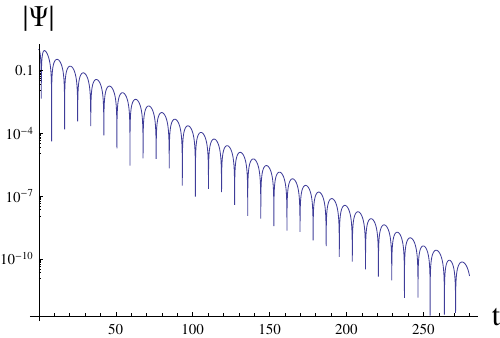}}
\caption{Time-domain profile  for $\ell=3$, $l_{cr}=1.137$. The Prony method gives $\omega = 0.59618 - 0.091002 i$ for the fundamental mode $n=0$.}\label{fig:L3}
\end{figure}

The difference between the two methods remains below $0.1\%$, whereas the deviation from the Schwarzschild limit 
\[
\omega_{\text{Schw}} = 0.3736715 - 0.0889625 i
\]
amounts to about $4\%$ for the damping rate and more than $1\%$ for the real oscillation frequency. In other words, for the lowest multipole $\ell=2$, the physical effect is at least one order of magnitude larger than the numerical uncertainty of the WKB approximation (see Figure~\ref{fig:L2}). Moreover, as $\ell$ increases, the WKB error decreases significantly, becoming two or more orders of magnitude smaller than the effect itself (see Figure~\ref {fig:L3}). Therefore, we conclude that the WKB method provides a reliable tool in this context. We can see that once the quantum correction is included, the real oscillation frequency and damping rate are decreased in comparison with their classical (Schwarzschild) limits.

Both the real oscillation frequency and the damping rate decrease as the quantum parameter $l_{\mathrm{cr}}$ increases, as can be seen for the fundamental mode and the first overtone across all multipole numbers $\ell = 2$–$4$ in Tables~I–VII. At the same time, for a fixed $\ell$, the first overtone exhibits significantly stronger deviations from its Schwarzschild limit than the fundamental mode. For instance, in the $\ell = 2$ case, the damping rate of the fundamental mode differs from the Schwarzschild value by approximately $4\%$, whereas the corresponding deviation for the first overtone exceeds $6\%$. The effect is even more pronounced for the real part of the frequency: while the shift of the fundamental mode is of the order of $1\%$, the first overtone shows a substantially larger deviation, reaching about $6\%$.

Notice that, by employing the correspondence between QNMs and grey-body factors \cite{Konoplya:2024lir,Lutfuoglu:2025ldc,Malik:2024cgb,Bolokhov:2024otn,Han:2025cal}, the recently obtained grey-body factors in \cite{Dubinsky:2025nxv} can be reproduced from the numerical data presented in this work.

A deviation at the level of a few percent in the complex QNM frequency is, in principle, within the target range of black-hole spectroscopy, but its detectability depends primarily on the ringdown signal-to-noise ratio (SNR), the adopted ringdown model (single mode vs.\ multimode/overtones), and correlations with the remnant mass and spin. Current LIGO--Virgo--KAGRA (LVK) analyses typically constrain QNM frequency parameters at the ${\cal O}(10\%)$ level (or weaker) for individual events, with the achievable precision strongly varying from event to event and depending on the ringdown start time and modeling assumptions \cite{LIGOScientific:2021sio,Ghosh:2021mrv,Cotesta:2022pci}. Within this landscape, a $\sim4\%$ shift in the damping rate and a $\sim1\%$ shift in the real oscillation frequency should be regarded as challenging for present detectors in single-event measurements, but potentially accessible for particularly loud ringdowns and, more robustly, through the combination (stacking) of multiple detections, for which statistical uncertainties typically scale as $\propto 1/\sqrt{N}$ at fixed systematics.

Future third-generation (3G) ground-based detectors are expected to deliver substantially higher ringdown SNRs and enable percent-level spectroscopy for a sizable population of stellar-mass binary black-hole mergers \cite{Bhagwat:2023jwv,Berti:2025hly}. Forecasts indicate that, for favorable sources, multiple QNM parameters (frequency and damping time/quality factor) can be measured with a few-percent accuracy, making shifts at the ${\cal O}(1\%$--$5\%)$ level observationally relevant in that era \cite{Bhagwat:2023jwv}. Therefore, while the present results mainly quantify the size of the effect at the waveform level, they also provide a concrete target for future ringdown tests in high-SNR events and 3G-era population analyses.

Thus, taking into account the recent advances in gravitational-wave observations, which have achieved very high signal-to-noise ratios \cite{KAGRA:2025oiz}, the estimated $4\%$ effect provides hope that such deviations may become observable. This raises the prospect of distinguishing classical black holes from their regular, quantum-corrected counterparts, even though the geometric modifications are confined primarily to a relatively small region near the event horizon.

Correspondence between QNMs and null geodesics. In the eikonal (geometric–optics) regime, when the multipole number $\ell$ is large, the dynamics of perturbations around a black hole is closely related to the properties of unstable circular null geodesics. In this limit, the wavefronts of the perturbing field propagate along null rays, and the effective potential governing quasinormal oscillations develops a sharp peak near the photon sphere. 
The real part of the quasinormal frequency is then determined by the angular velocity $\Omega_c$ of the null particle at this orbit, while the imaginary part is governed by the Lyapunov exponent $\lambda$, which characterizes the instability timescale of the corresponding geodesic trajectory \cite{Cardoso:2008bp}. 
Consequently, the fundamental quasinormal frequencies can be approximated as
\begin{equation}
\omega_{\ell n} \simeq \Omega_c\,\ell - i\left(n+\frac{1}{2}\right)\lambda,
\label{eikonal_qnm}
\end{equation}
where $n$ is the overtone number. This correspondence holds for a broad class of black holes with single–peaked effective potentials, providing an intuitive geometrical interpretation of the quasinormal spectrum in terms of the motion of photons on unstable circular orbits. Nevertheless, there are a number of deviations from this relation, which are described in \cite{Khanna:2016yow,Konoplya:2017wot,Bolokhov:2023dxq} and are usually related to gravitational or other non-minimally coupled perturbations. Thus, in some theories with higher-curvature corrections—such as Gauss–Bonnet, Lovelock, or Einstein-Weyl gravity — the centrifugal term that dominates in the eikonal regime does not retain its standard form \cite{Konoplya:2017lhs,Konoplya:2025afm}. Moreover, in asymptotically de Sitter spacetimes, an additional branch of QNMs emerges, corresponding to pure de Sitter modes modified by the presence of a black hole, for which the usual eikonal correspondence does not apply \cite{Konoplya:2025mvj,Konoplya:2022gjp}.  Such cases may signal the breakdown of the eikonal approximation. Here, in Table VIII we study the high $\ell$ regime of QNMs for the near-extreme Dymnikova black hole. One can easily compute the proper characteristics of the null geodesics for the Dymnikov black hole and compare them with the numerical data presented in Table VIII. We conclude that the correspondence holds for gravitational perturbations in our case. 

This can also be shown analytically. An unstable null circular geodesic at a distance $r=r_c$ satisfies the condition $$V'(r)\bigl|_{r_c}=0,$$ that leads to the following relation,
\begin{equation}\label{rccond}
2f_c-r_c f'_c=0.
\end{equation}
Then, the above relation is used to find the radius of the circular orbit $r_c$.

The angular velocity $$\Omega=d\varphi/dt=\dot{\varphi}/\dot{t}$$ can be found using the circular motion condition $\dot{r}=0$:
\begin{equation}\label{Omegac}
\Omega_c=\frac{\sqrt{f_c}}{r_c}.
\end{equation}
Finally, the Lyapunov exponent can be found as follows \cite{Cardoso:2008bp}, 
\begin{equation}
\lambda=\sqrt{\frac{f_c(2f_c-f''_cr_c^2)}{2r_c^2}}.
\end{equation}
By expanding the location of the maximum of the effective potential in powers of $1/\ell$, which governs the decay of gravitational perturbations, one immediately finds that it coincides with the radius of the circular null orbit $r_c$. Consequently, applying the eikonal limit of the first-order WKB approximation to the quasinormal frequency directly reproduces Eq.~\eqref{eikonal_qnm}.

\section{Conclusions}  \label{Concl}

In this work, we have analyzed the gravitational quasinormal spectrum of the Dymnikova black hole, a regular spacetime which can be interpreted either as a phenomenological model with a de Sitter core or as an effective geometry emerging from Asymptotically Safe gravity. Using the WKB method with Padé improvements, supported by time-domain integration, we have obtained the dominant quasinormal frequencies for axial gravitational perturbations. Our results show that the quantum correction parameter $l_{\rm cr}$ systematically modifies the spectrum: both the real oscillation frequencies and the damping rates decrease when the corrections are switched on, so that the ringdown signal becomes longer-lived compared to the Schwarzschild case. In the limit of large $l_{\rm cr}$, the Schwarzschild spectrum is smoothly recovered, confirming the consistency of the model. We also analyzed the regime of large multipole numbers and confirmed the correspondence between null geodesics and eikonal QNMs. 

It is important to emphasize that the present study has been mainly restricted to the fundamental mode and the first overtone for which the WKB approach is reliable. For higher overtones, however, the situation is expected to be qualitatively different. Since overtones are strongly localized near the event horizon, they should be especially sensitive to the local geometric deformations induced by quantum corrections \cite{Konoplya:2022pbc,Konoplya:2022hll}. Unfortunately, the WKB approximation is reliable only when the multipole number $\ell$ is larger than, or at most comparable to, the overtone number $n$, and it generally fails in the regime $n \gg \ell$. In this case, the imaginary part of the frequency becomes large and the corresponding turning points are widely separated in the complex plane, which undermines the accuracy of the WKB expansion. Although the WKB method is computationally efficient and highly effective for the lowest modes, it does not guarantee convergence order by order; instead, only asymptotic convergence can be expected. At the same time, the phenomenon of overtone outbursts occurs precisely in the regime where the overtone number appears to exceed the multipole number, so that the relevant physics is governed by the near-horizon geometry rather than by the potential peak dominated by the centrifugal, $\ell$-dependent term. However, even with the WKB method within its expected accuracy range, we see that the first overtone deviates from its Schwarzschild limit much stronger than the fundamental mode, indicating that there should be an outburst of overtones found for test fields in \cite{Konoplya:2023aph}. To investigate this regime, one would need to resort to more precise techniques such as the Frobenius (Leaver) method \cite{Leaver:1985ax}, which provides convergent series solutions and can capture the full spectrum beyond the eikonal approximation. At the same time, the metric, and consequently the master wave equation, does not have a rational form, so one has to transform or approximate the metric by some rational function for accurate calculations of higher overtones. Having in mind that the fundamental mode and possibly the first overtone are the only realistic candidates for detection in the near future experiments, we leave this analysis, as well as the inclusion of polar perturbations and possible extensions to higher dimensions, for future work.

\begin{acknowledgments}
A. S. is grateful to the Excellence project FoS UHK 2205/2025-2026 and other funds of the FoS UHK for the financial support. B. C. L. and J. R. are grateful to the Excellence project FoS UHK 2205/2025-2026 for the financial support.
\end{acknowledgments}

\bibliography{bibliography}

\begin{thebibliography}{115}%
\makeatletter
\providecommand \@ifxundefined [1]{%
 \@ifx{#1\undefined}
}%
\providecommand \@ifnum [1]{%
 \ifnum #1\expandafter \@firstoftwo
 \else \expandafter \@secondoftwo
 \fi
}%
\providecommand \@ifx [1]{%
 \ifx #1\expandafter \@firstoftwo
 \else \expandafter \@secondoftwo
 \fi
}%
\providecommand \natexlab [1]{#1}%
\providecommand \enquote  [1]{``#1''}%
\providecommand \bibnamefont  [1]{#1}%
\providecommand \bibfnamefont [1]{#1}%
\providecommand \citenamefont [1]{#1}%
\providecommand \href@noop [0]{\@secondoftwo}%
\providecommand \href [0]{\begingroup \@sanitize@url \@href}%
\providecommand \@href[1]{\@@startlink{#1}\@@href}%
\providecommand \@@href[1]{\endgroup#1\@@endlink}%
\providecommand \@sanitize@url [0]{\catcode `\\12\catcode `\$12\catcode `\&12\catcode `\#12\catcode `\^12\catcode `\_12\catcode `\%12\relax}%
\providecommand \@@startlink[1]{}%
\providecommand \@@endlink[0]{}%
\providecommand \url  [0]{\begingroup\@sanitize@url \@url }%
\providecommand \@url [1]{\endgroup\@href {#1}{\urlprefix }}%
\providecommand \urlprefix  [0]{URL }%
\providecommand \Eprint [0]{\href }%
\providecommand \doibase [0]{http://dx.doi.org/}%
\providecommand \selectlanguage [0]{\@gobble}%
\providecommand \bibinfo  [0]{\@secondoftwo}%
\providecommand \bibfield  [0]{\@secondoftwo}%
\providecommand \translation [1]{[#1]}%
\providecommand \BibitemOpen [0]{}%
\providecommand \bibitemStop [0]{}%
\providecommand \bibitemNoStop [0]{.\EOS\space}%
\providecommand \EOS [0]{\spacefactor3000\relax}%
\providecommand \BibitemShut  [1]{\csname bibitem#1\endcsname}%
\let\auto@bib@innerbib\@empty
\bibitem [{\citenamefont {Kokkotas}\ and\ \citenamefont {Schmidt}(1999)}]{Kokkotas:1999bd}%
  \BibitemOpen
  \bibfield  {author} {\bibinfo {author} {\bibfnamefont {K.~D.}\ \bibnamefont {Kokkotas}}\ and\ \bibinfo {author} {\bibfnamefont {B.~G.}\ \bibnamefont {Schmidt}},\ }\href {\doibase 10.12942/lrr-1999-2} {\bibfield  {journal} {\bibinfo  {journal} {Living Rev. Rel.}\ }\textbf {\bibinfo {volume} {2}},\ \bibinfo {pages} {2} (\bibinfo {year} {1999})},\ \Eprint {http://arxiv.org/abs/gr-qc/9909058} {arXiv:gr-qc/9909058} \BibitemShut {NoStop}%
\bibitem [{\citenamefont {Berti}\ \emph {et~al.}(2009)\citenamefont {Berti}, \citenamefont {Cardoso},\ and\ \citenamefont {Starinets}}]{Berti:2009kk}%
  \BibitemOpen
  \bibfield  {author} {\bibinfo {author} {\bibfnamefont {E.}~\bibnamefont {Berti}}, \bibinfo {author} {\bibfnamefont {V.}~\bibnamefont {Cardoso}}, \ and\ \bibinfo {author} {\bibfnamefont {A.~O.}\ \bibnamefont {Starinets}},\ }\href {\doibase 10.1088/0264-9381/26/16/163001} {\bibfield  {journal} {\bibinfo  {journal} {Class. Quant. Grav.}\ }\textbf {\bibinfo {volume} {26}},\ \bibinfo {pages} {163001} (\bibinfo {year} {2009})},\ \Eprint {http://arxiv.org/abs/0905.2975} {arXiv:0905.2975 [gr-qc]} \BibitemShut {NoStop}%
\bibitem [{\citenamefont {Konoplya}\ and\ \citenamefont {Zhidenko}(2011)}]{Konoplya:2011qq}%
  \BibitemOpen
  \bibfield  {author} {\bibinfo {author} {\bibfnamefont {R.~A.}\ \bibnamefont {Konoplya}}\ and\ \bibinfo {author} {\bibfnamefont {A.}~\bibnamefont {Zhidenko}},\ }\href {\doibase 10.1103/RevModPhys.83.793} {\bibfield  {journal} {\bibinfo  {journal} {Rev. Mod. Phys.}\ }\textbf {\bibinfo {volume} {83}},\ \bibinfo {pages} {793} (\bibinfo {year} {2011})},\ \Eprint {http://arxiv.org/abs/1102.4014} {arXiv:1102.4014 [gr-qc]} \BibitemShut {NoStop}%
\bibitem [{\citenamefont {Bolokhov}\ and\ \citenamefont {Skvortsova}(2025{\natexlab{a}})}]{Bolokhov:2025uxz}%
  \BibitemOpen
  \bibfield  {author} {\bibinfo {author} {\bibfnamefont {S.~V.}\ \bibnamefont {Bolokhov}}\ and\ \bibinfo {author} {\bibfnamefont {M.}~\bibnamefont {Skvortsova}},\ }\href {\doibase 10.1134/S0202289325700306} {\bibfield  {journal} {\bibinfo  {journal} {Grav. Cosmol.}\ }\textbf {\bibinfo {volume} {31}},\ \bibinfo {pages} {423} (\bibinfo {year} {2025}{\natexlab{a}})},\ \Eprint {http://arxiv.org/abs/2504.05014} {arXiv:2504.05014 [gr-qc]} \BibitemShut {NoStop}%
\bibitem [{\citenamefont {Abbott}\ \emph {et~al.}(2016{\natexlab{a}})\citenamefont {Abbott} \emph {et~al.}}]{LIGOScientific:2016aoc}%
  \BibitemOpen
  \bibfield  {author} {\bibinfo {author} {\bibfnamefont {B.~P.}\ \bibnamefont {Abbott}} \emph {et~al.} (\bibinfo {collaboration} {LIGO Scientific, Virgo}),\ }\href {\doibase 10.1103/PhysRevLett.116.061102} {\bibfield  {journal} {\bibinfo  {journal} {Phys. Rev. Lett.}\ }\textbf {\bibinfo {volume} {116}},\ \bibinfo {pages} {061102} (\bibinfo {year} {2016}{\natexlab{a}})},\ \Eprint {http://arxiv.org/abs/1602.03837} {arXiv:1602.03837 [gr-qc]} \BibitemShut {NoStop}%
\bibitem [{\citenamefont {Abbott}\ \emph {et~al.}(2017)\citenamefont {Abbott} \emph {et~al.}}]{LIGOScientific:2017vwq}%
  \BibitemOpen
  \bibfield  {author} {\bibinfo {author} {\bibfnamefont {B.~P.}\ \bibnamefont {Abbott}} \emph {et~al.} (\bibinfo {collaboration} {LIGO Scientific, Virgo}),\ }\href {\doibase 10.1103/PhysRevLett.119.161101} {\bibfield  {journal} {\bibinfo  {journal} {Phys. Rev. Lett.}\ }\textbf {\bibinfo {volume} {119}},\ \bibinfo {pages} {161101} (\bibinfo {year} {2017})},\ \Eprint {http://arxiv.org/abs/1710.05832} {arXiv:1710.05832 [gr-qc]} \BibitemShut {NoStop}%
\bibitem [{\citenamefont {Abbott}\ \emph {et~al.}(2020)\citenamefont {Abbott} \emph {et~al.}}]{LIGOScientific:2020zkf}%
  \BibitemOpen
  \bibfield  {author} {\bibinfo {author} {\bibfnamefont {R.}~\bibnamefont {Abbott}} \emph {et~al.} (\bibinfo {collaboration} {LIGO Scientific, Virgo}),\ }\href {\doibase 10.3847/2041-8213/ab960f} {\bibfield  {journal} {\bibinfo  {journal} {Astrophys. J. Lett.}\ }\textbf {\bibinfo {volume} {896}},\ \bibinfo {pages} {L44} (\bibinfo {year} {2020})},\ \Eprint {http://arxiv.org/abs/2006.12611} {arXiv:2006.12611 [astro-ph.HE]} \BibitemShut {NoStop}%
\bibitem [{\citenamefont {Abbott}\ \emph {et~al.}(2016{\natexlab{b}})\citenamefont {Abbott} \emph {et~al.}}]{KAGRA:2013rdx}%
  \BibitemOpen
  \bibfield  {author} {\bibinfo {author} {\bibfnamefont {B.~P.}\ \bibnamefont {Abbott}} \emph {et~al.} (\bibinfo {collaboration} {KAGRA, LIGO Scientific, Virgo}),\ }\href {\doibase 10.1007/s41114-020-00026-9} {\bibfield  {journal} {\bibinfo  {journal} {Living Rev. Rel.}\ }\textbf {\bibinfo {volume} {19}},\ \bibinfo {pages} {1} (\bibinfo {year} {2016}{\natexlab{b}})},\ \Eprint {http://arxiv.org/abs/1304.0670} {arXiv:1304.0670 [gr-qc]} \BibitemShut {NoStop}%
\bibitem [{\citenamefont {Turimov}\ \emph {et~al.}(2025)\citenamefont {Turimov}, \citenamefont {Usanov},\ and\ \citenamefont {Khamroev}}]{Turimov:2025tmf}%
  \BibitemOpen
  \bibfield  {author} {\bibinfo {author} {\bibfnamefont {B.}~\bibnamefont {Turimov}}, \bibinfo {author} {\bibfnamefont {S.}~\bibnamefont {Usanov}}, \ and\ \bibinfo {author} {\bibfnamefont {Y.}~\bibnamefont {Khamroev}},\ }\href {\doibase 10.1016/j.dark.2025.101876} {\bibfield  {journal} {\bibinfo  {journal} {Phys. Dark Univ.}\ }\textbf {\bibinfo {volume} {48}},\ \bibinfo {pages} {101876} (\bibinfo {year} {2025})},\ \Eprint {http://arxiv.org/abs/2502.11185} {arXiv:2502.11185 [gr-qc]} \BibitemShut {NoStop}%
\bibitem [{\citenamefont {{Derkach}}\ \emph {et~al.}(2025)\citenamefont {{Derkach}}, \citenamefont {{Trunk}}, \citenamefont {{Yusupov}},\ and\ \citenamefont {{Matrasulov}}}]{2025JPhA...58H5201D}%
  \BibitemOpen
  \bibfield  {author} {\bibinfo {author} {\bibfnamefont {V.~A.}\ \bibnamefont {{Derkach}}}, \bibinfo {author} {\bibfnamefont {C.}~\bibnamefont {{Trunk}}}, \bibinfo {author} {\bibfnamefont {J.~R.}\ \bibnamefont {{Yusupov}}}, \ and\ \bibinfo {author} {\bibfnamefont {D.~U.}\ \bibnamefont {{Matrasulov}}},\ }\href {\doibase 10.1088/1751-8121/adf787} {\bibfield  {journal} {\bibinfo  {journal} {J. Phys. A: Math. Gen.}\ }\textbf {\bibinfo {volume} {58}},\ \bibinfo {eid} {345201} (\bibinfo {year} {2025})},\ \Eprint {http://arxiv.org/abs/2410.10232} {arXiv:2410.10232 [math-ph]} \BibitemShut {NoStop}%
\bibitem [{\citenamefont {{Rakhmanov}}\ \emph {et~al.}(2025)\citenamefont {{Rakhmanov}}, \citenamefont {{Matchonov}}, \citenamefont {{Yusupov}}, \citenamefont {{Nasriddinov}},\ and\ \citenamefont {{Matrasulov}}}]{2025EPJB...98...35R}%
  \BibitemOpen
  \bibfield  {author} {\bibinfo {author} {\bibfnamefont {S.}~\bibnamefont {{Rakhmanov}}}, \bibinfo {author} {\bibfnamefont {K.}~\bibnamefont {{Matchonov}}}, \bibinfo {author} {\bibfnamefont {H.}~\bibnamefont {{Yusupov}}}, \bibinfo {author} {\bibfnamefont {K.}~\bibnamefont {{Nasriddinov}}}, \ and\ \bibinfo {author} {\bibfnamefont {D.}~\bibnamefont {{Matrasulov}}},\ }\href {\doibase 10.1140/epjb/s10051-025-00885-7} {\bibfield  {journal} {\bibinfo  {journal} {Eur. Phys. J. B}\ }\textbf {\bibinfo {volume} {98}},\ \bibinfo {eid} {35} (\bibinfo {year} {2025})},\ \Eprint {http://arxiv.org/abs/2504.03599} {arXiv:2504.03599 [cond-mat.mes-hall]} \BibitemShut {NoStop}%
\bibitem [{\citenamefont {Dymnikova}(1992)}]{Dymnikova:1992ux}%
  \BibitemOpen
  \bibfield  {author} {\bibinfo {author} {\bibfnamefont {I.}~\bibnamefont {Dymnikova}},\ }\href {\doibase 10.1007/BF00760226} {\bibfield  {journal} {\bibinfo  {journal} {Gen. Rel. Grav.}\ }\textbf {\bibinfo {volume} {24}},\ \bibinfo {pages} {235} (\bibinfo {year} {1992})}\BibitemShut {NoStop}%
\bibitem [{\citenamefont {Platania}(2019)}]{Platania:2019kyx}%
  \BibitemOpen
  \bibfield  {author} {\bibinfo {author} {\bibfnamefont {A.}~\bibnamefont {Platania}},\ }\href {\doibase 10.1140/epjc/s10052-019-6990-2} {\bibfield  {journal} {\bibinfo  {journal} {Eur. Phys. J. C}\ }\textbf {\bibinfo {volume} {79}},\ \bibinfo {pages} {470} (\bibinfo {year} {2019})},\ \Eprint {http://arxiv.org/abs/1903.10411} {arXiv:1903.10411 [gr-qc]} \BibitemShut {NoStop}%
\bibitem [{\citenamefont {Konoplya}\ and\ \citenamefont {Zhidenko}(2024{\natexlab{a}})}]{Konoplya:2024kih}%
  \BibitemOpen
  \bibfield  {author} {\bibinfo {author} {\bibfnamefont {R.~A.}\ \bibnamefont {Konoplya}}\ and\ \bibinfo {author} {\bibfnamefont {A.}~\bibnamefont {Zhidenko}},\ }\href {\doibase 10.1016/j.physletb.2024.138945} {\bibfield  {journal} {\bibinfo  {journal} {Phys. Lett. B}\ }\textbf {\bibinfo {volume} {856}},\ \bibinfo {pages} {138945} (\bibinfo {year} {2024}{\natexlab{a}})},\ \Eprint {http://arxiv.org/abs/2404.09063} {arXiv:2404.09063 [gr-qc]} \BibitemShut {NoStop}%
\bibitem [{\citenamefont {Mac{\^e}do}\ \emph {et~al.}(2024)\citenamefont {Mac{\^e}do}, \citenamefont {Furtado}, \citenamefont {Alencar},\ and\ \citenamefont {Landim}}]{Macedo:2024dqb}%
  \BibitemOpen
  \bibfield  {author} {\bibinfo {author} {\bibfnamefont {M.~H.}\ \bibnamefont {Mac{\^e}do}}, \bibinfo {author} {\bibfnamefont {J.}~\bibnamefont {Furtado}}, \bibinfo {author} {\bibfnamefont {G.}~\bibnamefont {Alencar}}, \ and\ \bibinfo {author} {\bibfnamefont {R.~R.}\ \bibnamefont {Landim}},\ }\href {\doibase 10.1016/j.aop.2024.169833} {\bibfield  {journal} {\bibinfo  {journal} {Annals Phys.}\ }\textbf {\bibinfo {volume} {471}},\ \bibinfo {pages} {169833} (\bibinfo {year} {2024})},\ \Eprint {http://arxiv.org/abs/2404.02818} {arXiv:2404.02818 [gr-qc]} \BibitemShut {NoStop}%
\bibitem [{\citenamefont {Konoplya}\ \emph {et~al.}(2023)\citenamefont {Konoplya}, \citenamefont {Stuchlik}, \citenamefont {Zhidenko},\ and\ \citenamefont {Zinhailo}}]{Konoplya:2023aph}%
  \BibitemOpen
  \bibfield  {author} {\bibinfo {author} {\bibfnamefont {R.~A.}\ \bibnamefont {Konoplya}}, \bibinfo {author} {\bibfnamefont {Z.}~\bibnamefont {Stuchlik}}, \bibinfo {author} {\bibfnamefont {A.}~\bibnamefont {Zhidenko}}, \ and\ \bibinfo {author} {\bibfnamefont {A.~F.}\ \bibnamefont {Zinhailo}},\ }\href {\doibase 10.1103/PhysRevD.107.104050} {\bibfield  {journal} {\bibinfo  {journal} {Phys. Rev. D}\ }\textbf {\bibinfo {volume} {107}},\ \bibinfo {pages} {104050} (\bibinfo {year} {2023})},\ \Eprint {http://arxiv.org/abs/2303.01987} {arXiv:2303.01987 [gr-qc]} \BibitemShut {NoStop}%
\bibitem [{\citenamefont {Dubinsky}(2026)}]{Dubinsky:2025nxv}%
  \BibitemOpen
  \bibfield  {author} {\bibinfo {author} {\bibfnamefont {A.}~\bibnamefont {Dubinsky}},\ }\href {\doibase 10.1016/j.aop.2025.170299} {\bibfield  {journal} {\bibinfo  {journal} {Annals Phys.}\ }\textbf {\bibinfo {volume} {485}},\ \bibinfo {pages} {170299} (\bibinfo {year} {2026})},\ \Eprint {http://arxiv.org/abs/2509.11017} {arXiv:2509.11017 [gr-qc]} \BibitemShut {NoStop}%
\bibitem [{\citenamefont {Konoplya}\ and\ \citenamefont {Zhidenko}(2020)}]{Konoplya:2020der}%
  \BibitemOpen
  \bibfield  {author} {\bibinfo {author} {\bibfnamefont {R.~A.}\ \bibnamefont {Konoplya}}\ and\ \bibinfo {author} {\bibfnamefont {A.}~\bibnamefont {Zhidenko}},\ }\href {\doibase 10.1016/j.physletb.2020.135607} {\bibfield  {journal} {\bibinfo  {journal} {Phys. Lett. B}\ }\textbf {\bibinfo {volume} {807}},\ \bibinfo {pages} {135607} (\bibinfo {year} {2020})},\ \Eprint {http://arxiv.org/abs/2005.02225} {arXiv:2005.02225 [gr-qc]} \BibitemShut {NoStop}%
\bibitem [{\citenamefont {Chen}\ \emph {et~al.}(2023)\citenamefont {Chen}, \citenamefont {Sathiyaraj}, \citenamefont {Hassanabadi}, \citenamefont {Yang}, \citenamefont {Long},\ and\ \citenamefont {Tu}}]{Chen:2023wkq}%
  \BibitemOpen
  \bibfield  {author} {\bibinfo {author} {\bibfnamefont {H.}~\bibnamefont {Chen}}, \bibinfo {author} {\bibfnamefont {T.}~\bibnamefont {Sathiyaraj}}, \bibinfo {author} {\bibfnamefont {H.}~\bibnamefont {Hassanabadi}}, \bibinfo {author} {\bibfnamefont {Y.}~\bibnamefont {Yang}}, \bibinfo {author} {\bibfnamefont {Z.~W.}\ \bibnamefont {Long}}, \ and\ \bibinfo {author} {\bibfnamefont {F.~Q.}\ \bibnamefont {Tu}},\ }\href {\doibase 10.1007/s12648-023-02734-8} {\bibfield  {journal} {\bibinfo  {journal} {Indian J. Phys.}\ }\textbf {\bibinfo {volume} {97}},\ \bibinfo {pages} {4481} (\bibinfo {year} {2023})}\BibitemShut {NoStop}%
\bibitem [{\citenamefont {Baruah}\ \emph {et~al.}(2023)\citenamefont {Baruah}, \citenamefont {{\"O}vg{\"u}n},\ and\ \citenamefont {Deshamukhya}}]{Baruah:2023rhd}%
  \BibitemOpen
  \bibfield  {author} {\bibinfo {author} {\bibfnamefont {A.}~\bibnamefont {Baruah}}, \bibinfo {author} {\bibfnamefont {A.}~\bibnamefont {{\"O}vg{\"u}n}}, \ and\ \bibinfo {author} {\bibfnamefont {A.}~\bibnamefont {Deshamukhya}},\ }\href {\doibase 10.1016/j.aop.2023.169393} {\bibfield  {journal} {\bibinfo  {journal} {Annals Phys.}\ }\textbf {\bibinfo {volume} {455}},\ \bibinfo {pages} {169393} (\bibinfo {year} {2023})},\ \Eprint {http://arxiv.org/abs/2304.07761} {arXiv:2304.07761 [gr-qc]} \BibitemShut {NoStop}%
\bibitem [{\citenamefont {Moreira}\ \emph {et~al.}(2023)\citenamefont {Moreira}, \citenamefont {Lima~Junior}, \citenamefont {Crispino},\ and\ \citenamefont {Herdeiro}}]{Moreira:2023cxy}%
  \BibitemOpen
  \bibfield  {author} {\bibinfo {author} {\bibfnamefont {Z.~S.}\ \bibnamefont {Moreira}}, \bibinfo {author} {\bibfnamefont {H.~C.~D.}\ \bibnamefont {Lima~Junior}}, \bibinfo {author} {\bibfnamefont {L.~C.~B.}\ \bibnamefont {Crispino}}, \ and\ \bibinfo {author} {\bibfnamefont {C.~A.~R.}\ \bibnamefont {Herdeiro}},\ }\href {\doibase 10.1103/PhysRevD.107.104016} {\bibfield  {journal} {\bibinfo  {journal} {Phys. Rev. D}\ }\textbf {\bibinfo {volume} {107}},\ \bibinfo {pages} {104016} (\bibinfo {year} {2023})},\ \Eprint {http://arxiv.org/abs/2302.14722} {arXiv:2302.14722 [gr-qc]} \BibitemShut {NoStop}%
\bibitem [{\citenamefont {Fu}\ \emph {et~al.}(2024)\citenamefont {Fu}, \citenamefont {Zhang}, \citenamefont {Liu}, \citenamefont {Kuang},\ and\ \citenamefont {Wu}}]{Fu:2023drp}%
  \BibitemOpen
  \bibfield  {author} {\bibinfo {author} {\bibfnamefont {G.}~\bibnamefont {Fu}}, \bibinfo {author} {\bibfnamefont {D.}~\bibnamefont {Zhang}}, \bibinfo {author} {\bibfnamefont {P.}~\bibnamefont {Liu}}, \bibinfo {author} {\bibfnamefont {X.-M.}\ \bibnamefont {Kuang}}, \ and\ \bibinfo {author} {\bibfnamefont {J.-P.}\ \bibnamefont {Wu}},\ }\href {\doibase 10.1103/PhysRevD.109.026010} {\bibfield  {journal} {\bibinfo  {journal} {Phys. Rev. D}\ }\textbf {\bibinfo {volume} {109}},\ \bibinfo {pages} {026010} (\bibinfo {year} {2024})},\ \Eprint {http://arxiv.org/abs/2301.08421} {arXiv:2301.08421 [gr-qc]} \BibitemShut {NoStop}%
\bibitem [{\citenamefont {Heidari}\ \emph {et~al.}(2023)\citenamefont {Heidari}, \citenamefont {Hassanabadi},\ and\ \citenamefont {Chen}}]{Heidari:2023ssx}%
  \BibitemOpen
  \bibfield  {author} {\bibinfo {author} {\bibfnamefont {N.}~\bibnamefont {Heidari}}, \bibinfo {author} {\bibfnamefont {H.}~\bibnamefont {Hassanabadi}}, \ and\ \bibinfo {author} {\bibfnamefont {H.}~\bibnamefont {Chen}},\ }\href {\doibase 10.1016/j.physletb.2023.137707} {\bibfield  {journal} {\bibinfo  {journal} {Phys. Lett. B}\ }\textbf {\bibinfo {volume} {838}},\ \bibinfo {pages} {137707} (\bibinfo {year} {2023})}\BibitemShut {NoStop}%
\bibitem [{\citenamefont {Skvortsova}(2025)}]{Skvortsova:2024msa}%
  \BibitemOpen
  \bibfield  {author} {\bibinfo {author} {\bibfnamefont {M.}~\bibnamefont {Skvortsova}},\ }\href {\doibase 10.1140/epjc/s10052-025-14589-w} {\bibfield  {journal} {\bibinfo  {journal} {Eur. Phys. J. C}\ }\textbf {\bibinfo {volume} {85}},\ \bibinfo {pages} {854} (\bibinfo {year} {2025})},\ \Eprint {http://arxiv.org/abs/2411.06007} {arXiv:2411.06007 [gr-qc]} \BibitemShut {NoStop}%
\bibitem [{\citenamefont {Kokkotas}\ \emph {et~al.}(2017)\citenamefont {Kokkotas}, \citenamefont {Konoplya},\ and\ \citenamefont {Zhidenko}}]{Kokkotas:2017zwt}%
  \BibitemOpen
  \bibfield  {author} {\bibinfo {author} {\bibfnamefont {K.}~\bibnamefont {Kokkotas}}, \bibinfo {author} {\bibfnamefont {R.~A.}\ \bibnamefont {Konoplya}}, \ and\ \bibinfo {author} {\bibfnamefont {A.}~\bibnamefont {Zhidenko}},\ }\href {\doibase 10.1103/PhysRevD.96.064007} {\bibfield  {journal} {\bibinfo  {journal} {Phys. Rev. D}\ }\textbf {\bibinfo {volume} {96}},\ \bibinfo {pages} {064007} (\bibinfo {year} {2017})},\ \Eprint {http://arxiv.org/abs/1705.09875} {arXiv:1705.09875 [gr-qc]} \BibitemShut {NoStop}%
\bibitem [{\citenamefont {Abdalla}\ \emph {et~al.}(2005)\citenamefont {Abdalla}, \citenamefont {Konoplya},\ and\ \citenamefont {Molina}}]{Abdalla:2005hu}%
  \BibitemOpen
  \bibfield  {author} {\bibinfo {author} {\bibfnamefont {E.}~\bibnamefont {Abdalla}}, \bibinfo {author} {\bibfnamefont {R.~A.}\ \bibnamefont {Konoplya}}, \ and\ \bibinfo {author} {\bibfnamefont {C.}~\bibnamefont {Molina}},\ }\href {\doibase 10.1103/PhysRevD.72.084006} {\bibfield  {journal} {\bibinfo  {journal} {Phys. Rev. D}\ }\textbf {\bibinfo {volume} {72}},\ \bibinfo {pages} {084006} (\bibinfo {year} {2005})},\ \Eprint {http://arxiv.org/abs/hep-th/0507100} {arXiv:hep-th/0507100} \BibitemShut {NoStop}%
\bibitem [{\citenamefont {Skvortsova}(2024{\natexlab{a}})}]{Skvortsova:2023zca}%
  \BibitemOpen
  \bibfield  {author} {\bibinfo {author} {\bibfnamefont {M.}~\bibnamefont {Skvortsova}},\ }\href {\doibase 10.1134/S0202289324010110} {\bibfield  {journal} {\bibinfo  {journal} {Grav. Cosmol.}\ }\textbf {\bibinfo {volume} {30}},\ \bibinfo {pages} {68} (\bibinfo {year} {2024}{\natexlab{a}})},\ \Eprint {http://arxiv.org/abs/2311.02729} {arXiv:2311.02729 [gr-qc]} \BibitemShut {NoStop}%
\bibitem [{\citenamefont {Bonanno}\ \emph {et~al.}(2025)\citenamefont {Bonanno}, \citenamefont {Konoplya}, \citenamefont {Oglialoro},\ and\ \citenamefont {Spina}}]{Bonanno:2025dry}%
  \BibitemOpen
  \bibfield  {author} {\bibinfo {author} {\bibfnamefont {A.~M.}\ \bibnamefont {Bonanno}}, \bibinfo {author} {\bibfnamefont {R.~A.}\ \bibnamefont {Konoplya}}, \bibinfo {author} {\bibfnamefont {G.}~\bibnamefont {Oglialoro}}, \ and\ \bibinfo {author} {\bibfnamefont {A.}~\bibnamefont {Spina}},\ }\href {\doibase 10.1088/1475-7516/2025/12/042} {\bibfield  {journal} {\bibinfo  {journal} {JCAP}\ }\textbf {\bibinfo {volume} {12}},\ \bibinfo {pages} {042} (\bibinfo {year} {2025})},\ \Eprint {http://arxiv.org/abs/2509.12469} {arXiv:2509.12469 [gr-qc]} \BibitemShut {NoStop}%
\bibitem [{\citenamefont {Bolokhov}\ \emph {et~al.}(2025)\citenamefont {Bolokhov}, \citenamefont {Bronnikov},\ and\ \citenamefont {Konoplya}}]{Bolokhov:2023ozp}%
  \BibitemOpen
  \bibfield  {author} {\bibinfo {author} {\bibfnamefont {S.}~\bibnamefont {Bolokhov}}, \bibinfo {author} {\bibfnamefont {K.}~\bibnamefont {Bronnikov}}, \ and\ \bibinfo {author} {\bibfnamefont {R.}~\bibnamefont {Konoplya}},\ }\href {\doibase 10.1002/prop.202400187} {\bibfield  {journal} {\bibinfo  {journal} {Fortsch. Phys.}\ }\textbf {\bibinfo {volume} {73}},\ \bibinfo {pages} {2400187} (\bibinfo {year} {2025})},\ \Eprint {http://arxiv.org/abs/2306.11083} {arXiv:2306.11083 [gr-qc]} \BibitemShut {NoStop}%
\bibitem [{\citenamefont {Bolokhov}\ and\ \citenamefont {Skvortsova}(2025{\natexlab{b}})}]{Bolokhov:2025lnt}%
  \BibitemOpen
  \bibfield  {author} {\bibinfo {author} {\bibfnamefont {S.~V.}\ \bibnamefont {Bolokhov}}\ and\ \bibinfo {author} {\bibfnamefont {M.}~\bibnamefont {Skvortsova}},\ }\href {\doibase 10.53941/ijgtp.2025.100003} {\bibfield  {journal} {\bibinfo  {journal} {Int. J. Grav. Theor. Phys.}\ }\textbf {\bibinfo {volume} {1}},\ \bibinfo {pages} {3} (\bibinfo {year} {2025}{\natexlab{b}})},\ \Eprint {http://arxiv.org/abs/2507.07196} {arXiv:2507.07196 [gr-qc]} \BibitemShut {NoStop}%
\bibitem [{\citenamefont {Malik}(2025{\natexlab{a}})}]{Malik:2024elk}%
  \BibitemOpen
  \bibfield  {author} {\bibinfo {author} {\bibfnamefont {Z.}~\bibnamefont {Malik}},\ }\href {\doibase 10.1007/s10773-024-05847-w} {\bibfield  {journal} {\bibinfo  {journal} {Int. J. Theor. Phys.}\ }\textbf {\bibinfo {volume} {64}},\ \bibinfo {pages} {30} (\bibinfo {year} {2025}{\natexlab{a}})}\BibitemShut {NoStop}%
\bibitem [{\citenamefont {Malik}(2025{\natexlab{b}})}]{Malik:2025dxn}%
  \BibitemOpen
  \bibfield  {author} {\bibinfo {author} {\bibfnamefont {Z.}~\bibnamefont {Malik}},\ }\href {\doibase 10.1007/s10773-025-06198-w} {\bibfield  {journal} {\bibinfo  {journal} {Int. J. Theor. Phys.}\ }\textbf {\bibinfo {volume} {64}},\ \bibinfo {pages} {314} (\bibinfo {year} {2025}{\natexlab{b}})},\ \Eprint {http://arxiv.org/abs/2508.19178} {arXiv:2508.19178 [gr-qc]} \BibitemShut {NoStop}%
\bibitem [{\citenamefont {Ashtekar}\ \emph {et~al.}(2018{\natexlab{a}})\citenamefont {Ashtekar}, \citenamefont {Olmedo},\ and\ \citenamefont {Singh}}]{Ashtekar:2018lag}%
  \BibitemOpen
  \bibfield  {author} {\bibinfo {author} {\bibfnamefont {A.}~\bibnamefont {Ashtekar}}, \bibinfo {author} {\bibfnamefont {J.}~\bibnamefont {Olmedo}}, \ and\ \bibinfo {author} {\bibfnamefont {P.}~\bibnamefont {Singh}},\ }\href {\doibase 10.1103/PhysRevLett.121.241301} {\bibfield  {journal} {\bibinfo  {journal} {Phys. Rev. Lett.}\ }\textbf {\bibinfo {volume} {121}},\ \bibinfo {pages} {241301} (\bibinfo {year} {2018}{\natexlab{a}})},\ \Eprint {http://arxiv.org/abs/1806.00648} {arXiv:1806.00648 [gr-qc]} \BibitemShut {NoStop}%
\bibitem [{\citenamefont {Ashtekar}\ \emph {et~al.}(2018{\natexlab{b}})\citenamefont {Ashtekar}, \citenamefont {Olmedo},\ and\ \citenamefont {Singh}}]{Ashtekar:2018cay}%
  \BibitemOpen
  \bibfield  {author} {\bibinfo {author} {\bibfnamefont {A.}~\bibnamefont {Ashtekar}}, \bibinfo {author} {\bibfnamefont {J.}~\bibnamefont {Olmedo}}, \ and\ \bibinfo {author} {\bibfnamefont {P.}~\bibnamefont {Singh}},\ }\href {\doibase 10.1103/PhysRevD.98.126003} {\bibfield  {journal} {\bibinfo  {journal} {Phys. Rev. D}\ }\textbf {\bibinfo {volume} {98}},\ \bibinfo {pages} {126003} (\bibinfo {year} {2018}{\natexlab{b}})},\ \Eprint {http://arxiv.org/abs/1806.02406} {arXiv:1806.02406 [gr-qc]} \BibitemShut {NoStop}%
\bibitem [{\citenamefont {Bouhmadi-L{\'o}pez}\ \emph {et~al.}(2020)\citenamefont {Bouhmadi-L{\'o}pez}, \citenamefont {Brahma}, \citenamefont {Chen}, \citenamefont {Chen},\ and\ \citenamefont {Yeom}}]{Bouhmadi-Lopez:2020oia}%
  \BibitemOpen
  \bibfield  {author} {\bibinfo {author} {\bibfnamefont {M.}~\bibnamefont {Bouhmadi-L{\'o}pez}}, \bibinfo {author} {\bibfnamefont {S.}~\bibnamefont {Brahma}}, \bibinfo {author} {\bibfnamefont {C.-Y.}\ \bibnamefont {Chen}}, \bibinfo {author} {\bibfnamefont {P.}~\bibnamefont {Chen}}, \ and\ \bibinfo {author} {\bibfnamefont {D.-h.}\ \bibnamefont {Yeom}},\ }\href {\doibase 10.1088/1475-7516/2020/07/066} {\bibfield  {journal} {\bibinfo  {journal} {JCAP}\ }\textbf {\bibinfo {volume} {07}},\ \bibinfo {pages} {066} (\bibinfo {year} {2020})},\ \Eprint {http://arxiv.org/abs/2004.13061} {arXiv:2004.13061 [gr-qc]} \BibitemShut {NoStop}%
\bibitem [{\citenamefont {Konoplya}\ and\ \citenamefont {Stashko}(2025{\natexlab{a}})}]{Konoplya:2024lch}%
  \BibitemOpen
  \bibfield  {author} {\bibinfo {author} {\bibfnamefont {R.~A.}\ \bibnamefont {Konoplya}}\ and\ \bibinfo {author} {\bibfnamefont {O.~S.}\ \bibnamefont {Stashko}},\ }\href {\doibase 10.1103/PhysRevD.111.104055} {\bibfield  {journal} {\bibinfo  {journal} {Phys. Rev. D}\ }\textbf {\bibinfo {volume} {111}},\ \bibinfo {pages} {104055} (\bibinfo {year} {2025}{\natexlab{a}})},\ \Eprint {http://arxiv.org/abs/2408.02578} {arXiv:2408.02578 [gr-qc]} \BibitemShut {NoStop}%
\bibitem [{\citenamefont {Bronnikov}\ \emph {et~al.}(2012)\citenamefont {Bronnikov}, \citenamefont {Konoplya},\ and\ \citenamefont {Zhidenko}}]{Bronnikov:2012ch}%
  \BibitemOpen
  \bibfield  {author} {\bibinfo {author} {\bibfnamefont {K.~A.}\ \bibnamefont {Bronnikov}}, \bibinfo {author} {\bibfnamefont {R.~A.}\ \bibnamefont {Konoplya}}, \ and\ \bibinfo {author} {\bibfnamefont {A.}~\bibnamefont {Zhidenko}},\ }\href {\doibase 10.1103/PhysRevD.86.024028} {\bibfield  {journal} {\bibinfo  {journal} {Phys. Rev. D}\ }\textbf {\bibinfo {volume} {86}},\ \bibinfo {pages} {024028} (\bibinfo {year} {2012})},\ \Eprint {http://arxiv.org/abs/1205.2224} {arXiv:1205.2224 [gr-qc]} \BibitemShut {NoStop}%
\bibitem [{\citenamefont {Chen}\ and\ \citenamefont {Chen}(2019)}]{Chen:2019iuo}%
  \BibitemOpen
  \bibfield  {author} {\bibinfo {author} {\bibfnamefont {C.-Y.}\ \bibnamefont {Chen}}\ and\ \bibinfo {author} {\bibfnamefont {P.}~\bibnamefont {Chen}},\ }\href {\doibase 10.1103/PhysRevD.99.104003} {\bibfield  {journal} {\bibinfo  {journal} {Phys. Rev. D}\ }\textbf {\bibinfo {volume} {99}},\ \bibinfo {pages} {104003} (\bibinfo {year} {2019})},\ \Eprint {http://arxiv.org/abs/1902.01678} {arXiv:1902.01678 [gr-qc]} \BibitemShut {NoStop}%
\bibitem [{\citenamefont {Konoplya}\ and\ \citenamefont {Stashko}(2025{\natexlab{b}})}]{Konoplya:2025hgp}%
  \BibitemOpen
  \bibfield  {author} {\bibinfo {author} {\bibfnamefont {R.~A.}\ \bibnamefont {Konoplya}}\ and\ \bibinfo {author} {\bibfnamefont {O.~S.}\ \bibnamefont {Stashko}},\ }\href {\doibase 10.1103/PhysRevD.111.084031} {\bibfield  {journal} {\bibinfo  {journal} {Phys. Rev. D}\ }\textbf {\bibinfo {volume} {111}},\ \bibinfo {pages} {084031} (\bibinfo {year} {2025}{\natexlab{b}})},\ \Eprint {http://arxiv.org/abs/2502.05689} {arXiv:2502.05689 [gr-qc]} \BibitemShut {NoStop}%
\bibitem [{\citenamefont {Bolokhov}\ and\ \citenamefont {Skvortsova}(2025{\natexlab{c}})}]{Bolokhov:2025egl}%
  \BibitemOpen
  \bibfield  {author} {\bibinfo {author} {\bibfnamefont {S.~V.}\ \bibnamefont {Bolokhov}}\ and\ \bibinfo {author} {\bibfnamefont {M.}~\bibnamefont {Skvortsova}},\ }\href@noop {} {\  (\bibinfo {year} {2025}{\natexlab{c}})},\ \Eprint {http://arxiv.org/abs/2508.19989} {arXiv:2508.19989 [gr-qc]} \BibitemShut {NoStop}%
\bibitem [{\citenamefont {Dymnikova}\ and\ \citenamefont {Galaktionov}(2005)}]{Dymnikova:2004qg}%
  \BibitemOpen
  \bibfield  {author} {\bibinfo {author} {\bibfnamefont {I.}~\bibnamefont {Dymnikova}}\ and\ \bibinfo {author} {\bibfnamefont {E.}~\bibnamefont {Galaktionov}},\ }\href {\doibase 10.1088/0264-9381/22/12/003} {\bibfield  {journal} {\bibinfo  {journal} {Class. Quant. Grav.}\ }\textbf {\bibinfo {volume} {22}},\ \bibinfo {pages} {2331} (\bibinfo {year} {2005})},\ \Eprint {http://arxiv.org/abs/gr-qc/0409049} {arXiv:gr-qc/0409049} \BibitemShut {NoStop}%
\bibitem [{\citenamefont {Berti}\ \emph {et~al.}(2003)\citenamefont {Berti}, \citenamefont {Kokkotas},\ and\ \citenamefont {Papantonopoulos}}]{Berti:2003yr}%
  \BibitemOpen
  \bibfield  {author} {\bibinfo {author} {\bibfnamefont {E.}~\bibnamefont {Berti}}, \bibinfo {author} {\bibfnamefont {K.~D.}\ \bibnamefont {Kokkotas}}, \ and\ \bibinfo {author} {\bibfnamefont {E.}~\bibnamefont {Papantonopoulos}},\ }\href {\doibase 10.1103/PhysRevD.68.064020} {\bibfield  {journal} {\bibinfo  {journal} {Phys. Rev. D}\ }\textbf {\bibinfo {volume} {68}},\ \bibinfo {pages} {064020} (\bibinfo {year} {2003})},\ \Eprint {http://arxiv.org/abs/gr-qc/0306106} {arXiv:gr-qc/0306106} \BibitemShut {NoStop}%
\bibitem [{\citenamefont {Kokkotas}(1993)}]{Kokkotas:1993ef}%
  \BibitemOpen
  \bibfield  {author} {\bibinfo {author} {\bibfnamefont {K.~D.}\ \bibnamefont {Kokkotas}},\ }\href {\doibase 10.1007/BF02822861} {\bibfield  {journal} {\bibinfo  {journal} {Nuovo Cim. B}\ }\textbf {\bibinfo {volume} {108}},\ \bibinfo {pages} {991} (\bibinfo {year} {1993})}\BibitemShut {NoStop}%
\bibitem [{\citenamefont {Konoplya}\ and\ \citenamefont {Zhidenko}(2007)}]{Konoplya:2006ar}%
  \BibitemOpen
  \bibfield  {author} {\bibinfo {author} {\bibfnamefont {R.~A.}\ \bibnamefont {Konoplya}}\ and\ \bibinfo {author} {\bibfnamefont {A.}~\bibnamefont {Zhidenko}},\ }\href {\doibase 10.1016/j.physletb.2007.03.018} {\bibfield  {journal} {\bibinfo  {journal} {Phys. Lett. B}\ }\textbf {\bibinfo {volume} {648}},\ \bibinfo {pages} {236} (\bibinfo {year} {2007})},\ \Eprint {http://arxiv.org/abs/hep-th/0611226} {arXiv:hep-th/0611226} \BibitemShut {NoStop}%
\bibitem [{\citenamefont {Schutz}\ and\ \citenamefont {Will}(1985)}]{Schutz:1985km}%
  \BibitemOpen
  \bibfield  {author} {\bibinfo {author} {\bibfnamefont {B.~F.}\ \bibnamefont {Schutz}}\ and\ \bibinfo {author} {\bibfnamefont {C.~M.}\ \bibnamefont {Will}},\ }\href {\doibase 10.1086/184453} {\bibfield  {journal} {\bibinfo  {journal} {Astrophys. J. Lett.}\ }\textbf {\bibinfo {volume} {291}},\ \bibinfo {pages} {L33} (\bibinfo {year} {1985})}\BibitemShut {NoStop}%
\bibitem [{\citenamefont {Iyer}\ and\ \citenamefont {Will}(1987)}]{Iyer:1986np}%
  \BibitemOpen
  \bibfield  {author} {\bibinfo {author} {\bibfnamefont {S.}~\bibnamefont {Iyer}}\ and\ \bibinfo {author} {\bibfnamefont {C.~M.}\ \bibnamefont {Will}},\ }\href {\doibase 10.1103/PhysRevD.35.3621} {\bibfield  {journal} {\bibinfo  {journal} {Phys. Rev. D}\ }\textbf {\bibinfo {volume} {35}},\ \bibinfo {pages} {3621} (\bibinfo {year} {1987})}\BibitemShut {NoStop}%
\bibitem [{\citenamefont {Konoplya}(2003)}]{Konoplya:2003ii}%
  \BibitemOpen
  \bibfield  {author} {\bibinfo {author} {\bibfnamefont {R.~A.}\ \bibnamefont {Konoplya}},\ }\href {\doibase 10.1103/PhysRevD.68.024018} {\bibfield  {journal} {\bibinfo  {journal} {Phys. Rev. D}\ }\textbf {\bibinfo {volume} {68}},\ \bibinfo {pages} {024018} (\bibinfo {year} {2003})},\ \Eprint {http://arxiv.org/abs/gr-qc/0303052} {arXiv:gr-qc/0303052} \BibitemShut {NoStop}%
\bibitem [{\citenamefont {Matyjasek}\ and\ \citenamefont {Opala}(2017)}]{Matyjasek:2017psv}%
  \BibitemOpen
  \bibfield  {author} {\bibinfo {author} {\bibfnamefont {J.}~\bibnamefont {Matyjasek}}\ and\ \bibinfo {author} {\bibfnamefont {M.}~\bibnamefont {Opala}},\ }\href {\doibase 10.1103/PhysRevD.96.024011} {\bibfield  {journal} {\bibinfo  {journal} {Phys. Rev. D}\ }\textbf {\bibinfo {volume} {96}},\ \bibinfo {pages} {024011} (\bibinfo {year} {2017})},\ \Eprint {http://arxiv.org/abs/1704.00361} {arXiv:1704.00361 [gr-qc]} \BibitemShut {NoStop}%
\bibitem [{\citenamefont {Konoplya}\ \emph {et~al.}(2019)\citenamefont {Konoplya}, \citenamefont {Zhidenko},\ and\ \citenamefont {Zinhailo}}]{Konoplya:2019hlu}%
  \BibitemOpen
  \bibfield  {author} {\bibinfo {author} {\bibfnamefont {R.~A.}\ \bibnamefont {Konoplya}}, \bibinfo {author} {\bibfnamefont {A.}~\bibnamefont {Zhidenko}}, \ and\ \bibinfo {author} {\bibfnamefont {A.~F.}\ \bibnamefont {Zinhailo}},\ }\href {\doibase 10.1088/1361-6382/ab2e25} {\bibfield  {journal} {\bibinfo  {journal} {Class. Quant. Grav.}\ }\textbf {\bibinfo {volume} {36}},\ \bibinfo {pages} {155002} (\bibinfo {year} {2019})},\ \Eprint {http://arxiv.org/abs/1904.10333} {arXiv:1904.10333 [gr-qc]} \BibitemShut {NoStop}%
\bibitem [{\citenamefont {Matyjasek}(2021)}]{Matyjasek:2021xfg}%
  \BibitemOpen
  \bibfield  {author} {\bibinfo {author} {\bibfnamefont {J.}~\bibnamefont {Matyjasek}},\ }\href {\doibase 10.1103/PhysRevD.104.084066} {\bibfield  {journal} {\bibinfo  {journal} {Phys. Rev. D}\ }\textbf {\bibinfo {volume} {104}},\ \bibinfo {pages} {084066} (\bibinfo {year} {2021})},\ \Eprint {http://arxiv.org/abs/2107.04815} {arXiv:2107.04815 [gr-qc]} \BibitemShut {NoStop}%
\bibitem [{\citenamefont {Pedrotti}\ and\ \citenamefont {Calz{\`a}}(2025)}]{Pedrotti:2025idg}%
  \BibitemOpen
  \bibfield  {author} {\bibinfo {author} {\bibfnamefont {D.}~\bibnamefont {Pedrotti}}\ and\ \bibinfo {author} {\bibfnamefont {M.}~\bibnamefont {Calz{\`a}}},\ }\href {\doibase 10.1103/1q35-mjjz} {\bibfield  {journal} {\bibinfo  {journal} {Phys. Rev. D}\ }\textbf {\bibinfo {volume} {111}},\ \bibinfo {pages} {124056} (\bibinfo {year} {2025})},\ \Eprint {http://arxiv.org/abs/2504.01909} {arXiv:2504.01909 [gr-qc]} \BibitemShut {NoStop}%
\bibitem [{\citenamefont {Konoplya}\ and\ \citenamefont {Zinhailo}(2020)}]{Konoplya:2020cbv}%
  \BibitemOpen
  \bibfield  {author} {\bibinfo {author} {\bibfnamefont {R.~A.}\ \bibnamefont {Konoplya}}\ and\ \bibinfo {author} {\bibfnamefont {A.~F.}\ \bibnamefont {Zinhailo}},\ }\href {\doibase 10.1016/j.physletb.2020.135793} {\bibfield  {journal} {\bibinfo  {journal} {Phys. Lett. B}\ }\textbf {\bibinfo {volume} {810}},\ \bibinfo {pages} {135793} (\bibinfo {year} {2020})},\ \Eprint {http://arxiv.org/abs/2004.02248} {arXiv:2004.02248 [gr-qc]} \BibitemShut {NoStop}%
\bibitem [{\citenamefont {Mahdavian~Yekta}\ \emph {et~al.}(2021)\citenamefont {Mahdavian~Yekta}, \citenamefont {Karimabadi},\ and\ \citenamefont {Alavi}}]{MahdavianYekta:2019pol}%
  \BibitemOpen
  \bibfield  {author} {\bibinfo {author} {\bibfnamefont {D.}~\bibnamefont {Mahdavian~Yekta}}, \bibinfo {author} {\bibfnamefont {M.}~\bibnamefont {Karimabadi}}, \ and\ \bibinfo {author} {\bibfnamefont {S.~A.}\ \bibnamefont {Alavi}},\ }\href {\doibase 10.1016/j.aop.2021.168603} {\bibfield  {journal} {\bibinfo  {journal} {Annals Phys.}\ }\textbf {\bibinfo {volume} {434}},\ \bibinfo {pages} {168603} (\bibinfo {year} {2021})},\ \Eprint {http://arxiv.org/abs/1912.12017} {arXiv:1912.12017 [hep-th]} \BibitemShut {NoStop}%
\bibitem [{\citenamefont {Zhang}\ \emph {et~al.}(2025)\citenamefont {Zhang}, \citenamefont {Zou}, \citenamefont {Zhang}, \citenamefont {Zhang},\ and\ \citenamefont {Yue}}]{Zhang:2025xqt}%
  \BibitemOpen
  \bibfield  {author} {\bibinfo {author} {\bibfnamefont {X.}~\bibnamefont {Zhang}}, \bibinfo {author} {\bibfnamefont {D.-C.}\ \bibnamefont {Zou}}, \bibinfo {author} {\bibfnamefont {C.-M.}\ \bibnamefont {Zhang}}, \bibinfo {author} {\bibfnamefont {M.}~\bibnamefont {Zhang}}, \ and\ \bibinfo {author} {\bibfnamefont {R.-H.}\ \bibnamefont {Yue}},\ }\href {\doibase 10.1088/1674-1137/ade661} {\bibfield  {journal} {\bibinfo  {journal} {Chin. Phys.}\ }\textbf {\bibinfo {volume} {49}},\ \bibinfo {pages} {105109} (\bibinfo {year} {2025})},\ \Eprint {http://arxiv.org/abs/2508.17736} {arXiv:2508.17736 [gr-qc]} \BibitemShut {NoStop}%
\bibitem [{\citenamefont {Antonelli}\ \emph {et~al.}(2025)\citenamefont {Antonelli}, \citenamefont {Giusti}, \citenamefont {Casadio},\ and\ \citenamefont {Heisenberg}}]{Antonelli:2025yol}%
  \BibitemOpen
  \bibfield  {author} {\bibinfo {author} {\bibfnamefont {T.}~\bibnamefont {Antonelli}}, \bibinfo {author} {\bibfnamefont {A.}~\bibnamefont {Giusti}}, \bibinfo {author} {\bibfnamefont {R.}~\bibnamefont {Casadio}}, \ and\ \bibinfo {author} {\bibfnamefont {L.}~\bibnamefont {Heisenberg}},\ }\href@noop {} {\  (\bibinfo {year} {2025})},\ \Eprint {http://arxiv.org/abs/2505.08415} {arXiv:2505.08415 [gr-qc]} \BibitemShut {NoStop}%
\bibitem [{\citenamefont {L{\"u}tf{\"u}o{\u{g}}lu}(2025{\natexlab{a}})}]{Lutfuoglu:2025ljm}%
  \BibitemOpen
  \bibfield  {author} {\bibinfo {author} {\bibfnamefont {B.~C.}\ \bibnamefont {L{\"u}tf{\"u}o{\u{g}}lu}},\ }\href {\doibase 10.1140/epjc/s10052-025-14380-x} {\bibfield  {journal} {\bibinfo  {journal} {Eur. Phys. J. C}\ }\textbf {\bibinfo {volume} {85}},\ \bibinfo {pages} {630} (\bibinfo {year} {2025}{\natexlab{a}})},\ \Eprint {http://arxiv.org/abs/2504.18482} {arXiv:2504.18482 [gr-qc]} \BibitemShut {NoStop}%
\bibitem [{\citenamefont {Hamil}\ and\ \citenamefont {L{\"u}tf{\"u}o{\u{g}}lu}(2025)}]{Hamil:2025cms}%
  \BibitemOpen
  \bibfield  {author} {\bibinfo {author} {\bibfnamefont {B.}~\bibnamefont {Hamil}}\ and\ \bibinfo {author} {\bibfnamefont {B.~C.}\ \bibnamefont {L{\"u}tf{\"u}o{\u{g}}lu}},\ }\href {\doibase 10.1007/s10714-025-03478-y} {\bibfield  {journal} {\bibinfo  {journal} {Gen. Rel. Grav.}\ }\textbf {\bibinfo {volume} {57}},\ \bibinfo {pages} {140} (\bibinfo {year} {2025})},\ \Eprint {http://arxiv.org/abs/2503.17474} {arXiv:2503.17474 [gr-qc]} \BibitemShut {NoStop}%
\bibitem [{\citenamefont {Miyachi}\ \emph {et~al.}(2025)\citenamefont {Miyachi}, \citenamefont {Namba}, \citenamefont {Omiya},\ and\ \citenamefont {Oshita}}]{Miyachi:2025ptm}%
  \BibitemOpen
  \bibfield  {author} {\bibinfo {author} {\bibfnamefont {T.}~\bibnamefont {Miyachi}}, \bibinfo {author} {\bibfnamefont {R.}~\bibnamefont {Namba}}, \bibinfo {author} {\bibfnamefont {H.}~\bibnamefont {Omiya}}, \ and\ \bibinfo {author} {\bibfnamefont {N.}~\bibnamefont {Oshita}},\ }\href {\doibase 10.1103/1gmr-9f1g} {\bibfield  {journal} {\bibinfo  {journal} {Phys. Rev. D}\ }\textbf {\bibinfo {volume} {111}},\ \bibinfo {pages} {124045} (\bibinfo {year} {2025})},\ \Eprint {http://arxiv.org/abs/2503.17245} {arXiv:2503.17245 [hep-th]} \BibitemShut {NoStop}%
\bibitem [{\citenamefont {Konoplya}\ and\ \citenamefont {Abdalla}(2005)}]{Konoplya:2005sy}%
  \BibitemOpen
  \bibfield  {author} {\bibinfo {author} {\bibfnamefont {R.~A.}\ \bibnamefont {Konoplya}}\ and\ \bibinfo {author} {\bibfnamefont {E.}~\bibnamefont {Abdalla}},\ }\href {\doibase 10.1103/PhysRevD.71.084015} {\bibfield  {journal} {\bibinfo  {journal} {Phys. Rev. D}\ }\textbf {\bibinfo {volume} {71}},\ \bibinfo {pages} {084015} (\bibinfo {year} {2005})},\ \Eprint {http://arxiv.org/abs/hep-th/0503029} {arXiv:hep-th/0503029} \BibitemShut {NoStop}%
\bibitem [{\citenamefont {Konoplya}(2021)}]{Konoplya:2021ube}%
  \BibitemOpen
  \bibfield  {author} {\bibinfo {author} {\bibfnamefont {R.~A.}\ \bibnamefont {Konoplya}},\ }\href {\doibase 10.1016/j.physletb.2021.136734} {\bibfield  {journal} {\bibinfo  {journal} {Phys. Lett. B}\ }\textbf {\bibinfo {volume} {823}},\ \bibinfo {pages} {136734} (\bibinfo {year} {2021})},\ \Eprint {http://arxiv.org/abs/2109.01640} {arXiv:2109.01640 [gr-qc]} \BibitemShut {NoStop}%
\bibitem [{\citenamefont {Konoplya}\ and\ \citenamefont {Zhidenko}(2023)}]{Konoplya:2023moy}%
  \BibitemOpen
  \bibfield  {author} {\bibinfo {author} {\bibfnamefont {R.~A.}\ \bibnamefont {Konoplya}}\ and\ \bibinfo {author} {\bibfnamefont {A.}~\bibnamefont {Zhidenko}},\ }\href {\doibase 10.1088/1361-6382/ad0a52} {\bibfield  {journal} {\bibinfo  {journal} {Class. Quant. Grav.}\ }\textbf {\bibinfo {volume} {40}},\ \bibinfo {pages} {245005} (\bibinfo {year} {2023})},\ \Eprint {http://arxiv.org/abs/2309.02560} {arXiv:2309.02560 [gr-qc]} \BibitemShut {NoStop}%
\bibitem [{\citenamefont {Dubinsky}(2025{\natexlab{a}})}]{Dubinsky:2025fwv}%
  \BibitemOpen
  \bibfield  {author} {\bibinfo {author} {\bibfnamefont {A.}~\bibnamefont {Dubinsky}},\ }\href {\doibase 10.53941/ijgtp.2025.100002} {\bibfield  {journal} {\bibinfo  {journal} {Int. J. Grav. Theor. Phys.}\ }\textbf {\bibinfo {volume} {1}},\ \bibinfo {pages} {2} (\bibinfo {year} {2025}{\natexlab{a}})},\ \Eprint {http://arxiv.org/abs/2507.00256} {arXiv:2507.00256 [gr-qc]} \BibitemShut {NoStop}%
\bibitem [{\citenamefont {Skvortsova}(2024{\natexlab{b}})}]{Skvortsova:2023zmj}%
  \BibitemOpen
  \bibfield  {author} {\bibinfo {author} {\bibfnamefont {M.}~\bibnamefont {Skvortsova}},\ }\href {\doibase 10.1002/prop.202400036} {\bibfield  {journal} {\bibinfo  {journal} {Fortsch. Phys.}\ }\textbf {\bibinfo {volume} {72}},\ \bibinfo {pages} {2400036} (\bibinfo {year} {2024}{\natexlab{b}})},\ \Eprint {http://arxiv.org/abs/2311.11650} {arXiv:2311.11650 [gr-qc]} \BibitemShut {NoStop}%
\bibitem [{\citenamefont {Bronnikov}\ and\ \citenamefont {Konoplya}(2020)}]{Bronnikov:2019sbx}%
  \BibitemOpen
  \bibfield  {author} {\bibinfo {author} {\bibfnamefont {K.~A.}\ \bibnamefont {Bronnikov}}\ and\ \bibinfo {author} {\bibfnamefont {R.~A.}\ \bibnamefont {Konoplya}},\ }\href {\doibase 10.1103/PhysRevD.101.064004} {\bibfield  {journal} {\bibinfo  {journal} {Phys. Rev. D}\ }\textbf {\bibinfo {volume} {101}},\ \bibinfo {pages} {064004} (\bibinfo {year} {2020})},\ \Eprint {http://arxiv.org/abs/1912.05315} {arXiv:1912.05315 [gr-qc]} \BibitemShut {NoStop}%
\bibitem [{\citenamefont {Konoplya}(2020)}]{Konoplya:2019xmn}%
  \BibitemOpen
  \bibfield  {author} {\bibinfo {author} {\bibfnamefont {R.~A.}\ \bibnamefont {Konoplya}},\ }\href {\doibase 10.1016/j.physletb.2020.135363} {\bibfield  {journal} {\bibinfo  {journal} {Phys. Lett. B}\ }\textbf {\bibinfo {volume} {804}},\ \bibinfo {pages} {135363} (\bibinfo {year} {2020})},\ \Eprint {http://arxiv.org/abs/1912.10582} {arXiv:1912.10582 [gr-qc]} \BibitemShut {NoStop}%
\bibitem [{\citenamefont {Bolokhov}(2024{\natexlab{a}})}]{Bolokhov:2024ixe}%
  \BibitemOpen
  \bibfield  {author} {\bibinfo {author} {\bibfnamefont {S.~V.}\ \bibnamefont {Bolokhov}},\ }\href {\doibase 10.1140/epjc/s10052-024-12990-5} {\bibfield  {journal} {\bibinfo  {journal} {Eur. Phys. J. C}\ }\textbf {\bibinfo {volume} {84}},\ \bibinfo {pages} {634} (\bibinfo {year} {2024}{\natexlab{a}})},\ \Eprint {http://arxiv.org/abs/2404.09364} {arXiv:2404.09364 [gr-qc]} \BibitemShut {NoStop}%
\bibitem [{\citenamefont {Dubinsky}(2025{\natexlab{b}})}]{Dubinsky:2025azv}%
  \BibitemOpen
  \bibfield  {author} {\bibinfo {author} {\bibfnamefont {A.}~\bibnamefont {Dubinsky}},\ }\href {\doibase 10.1007/s10773-025-06053-y} {\bibfield  {journal} {\bibinfo  {journal} {Int. J. Theor. Phys.}\ }\textbf {\bibinfo {volume} {64}},\ \bibinfo {pages} {203} (\bibinfo {year} {2025}{\natexlab{b}})},\ \Eprint {http://arxiv.org/abs/2405.13552} {arXiv:2405.13552 [gr-qc]} \BibitemShut {NoStop}%
\bibitem [{\citenamefont {Skvortsova}(2024{\natexlab{c}})}]{Skvortsova:2024atk}%
  \BibitemOpen
  \bibfield  {author} {\bibinfo {author} {\bibfnamefont {M.}~\bibnamefont {Skvortsova}},\ }\href {\doibase 10.1002/prop.202400132} {\bibfield  {journal} {\bibinfo  {journal} {Fortsch. Phys.}\ }\textbf {\bibinfo {volume} {72}},\ \bibinfo {pages} {2400132} (\bibinfo {year} {2024}{\natexlab{c}})},\ \Eprint {http://arxiv.org/abs/2405.06390} {arXiv:2405.06390 [gr-qc]} \BibitemShut {NoStop}%
\bibitem [{\citenamefont {Dubinsky}\ and\ \citenamefont {Zinhailo}(2024)}]{Dubinsky:2024hmn}%
  \BibitemOpen
  \bibfield  {author} {\bibinfo {author} {\bibfnamefont {A.}~\bibnamefont {Dubinsky}}\ and\ \bibinfo {author} {\bibfnamefont {A.}~\bibnamefont {Zinhailo}},\ }\href {\doibase 10.1140/epjc/s10052-024-13206-6} {\bibfield  {journal} {\bibinfo  {journal} {Eur. Phys. J. C}\ }\textbf {\bibinfo {volume} {84}},\ \bibinfo {pages} {847} (\bibinfo {year} {2024})},\ \Eprint {http://arxiv.org/abs/2404.01834} {arXiv:2404.01834 [gr-qc]} \BibitemShut {NoStop}%
\bibitem [{\citenamefont {Malik}(2024{\natexlab{a}})}]{Malik:2023bxc}%
  \BibitemOpen
  \bibfield  {author} {\bibinfo {author} {\bibfnamefont {Z.}~\bibnamefont {Malik}},\ }\href {\doibase 10.1007/s10773-024-05737-1} {\bibfield  {journal} {\bibinfo  {journal} {Int. J. Theor. Phys.}\ }\textbf {\bibinfo {volume} {63}},\ \bibinfo {pages} {199} (\bibinfo {year} {2024}{\natexlab{a}})},\ \Eprint {http://arxiv.org/abs/2308.10412} {arXiv:2308.10412 [gr-qc]} \BibitemShut {NoStop}%
\bibitem [{\citenamefont {Churilova}\ \emph {et~al.}(2021)\citenamefont {Churilova}, \citenamefont {Konoplya}, \citenamefont {Stuchlik},\ and\ \citenamefont {Zhidenko}}]{Churilova:2021tgn}%
  \BibitemOpen
  \bibfield  {author} {\bibinfo {author} {\bibfnamefont {M.~S.}\ \bibnamefont {Churilova}}, \bibinfo {author} {\bibfnamefont {R.~A.}\ \bibnamefont {Konoplya}}, \bibinfo {author} {\bibfnamefont {Z.}~\bibnamefont {Stuchlik}}, \ and\ \bibinfo {author} {\bibfnamefont {A.}~\bibnamefont {Zhidenko}},\ }\href {\doibase 10.1088/1475-7516/2021/10/010} {\bibfield  {journal} {\bibinfo  {journal} {JCAP}\ }\textbf {\bibinfo {volume} {10}},\ \bibinfo {pages} {010} (\bibinfo {year} {2021})},\ \Eprint {http://arxiv.org/abs/2107.05977} {arXiv:2107.05977 [gr-qc]} \BibitemShut {NoStop}%
\bibitem [{\citenamefont {L{\"u}tf{\"u}o{\u{g}}lu}(2025{\natexlab{b}})}]{Lutfuoglu:2025hjy}%
  \BibitemOpen
  \bibfield  {author} {\bibinfo {author} {\bibfnamefont {B.~C.}\ \bibnamefont {L{\"u}tf{\"u}o{\u{g}}lu}},\ }\href {\doibase 10.1140/epjc/s10052-025-14210-0} {\bibfield  {journal} {\bibinfo  {journal} {Eur. Phys. J. C}\ }\textbf {\bibinfo {volume} {85}},\ \bibinfo {pages} {486} (\bibinfo {year} {2025}{\natexlab{b}})},\ \Eprint {http://arxiv.org/abs/2503.16087} {arXiv:2503.16087 [gr-qc]} \BibitemShut {NoStop}%
\bibitem [{\citenamefont {Dubinsky}(2025{\natexlab{c}})}]{Dubinsky:2024rvf}%
  \BibitemOpen
  \bibfield  {author} {\bibinfo {author} {\bibfnamefont {A.}~\bibnamefont {Dubinsky}},\ }\href {\doibase 10.1016/j.physletb.2025.139251} {\bibfield  {journal} {\bibinfo  {journal} {Phys. Lett. B}\ }\textbf {\bibinfo {volume} {861}},\ \bibinfo {pages} {139251} (\bibinfo {year} {2025}{\natexlab{c}})},\ \Eprint {http://arxiv.org/abs/2409.16569} {arXiv:2409.16569 [gr-qc]} \BibitemShut {NoStop}%
\bibitem [{\citenamefont {Malik}(2025{\natexlab{c}})}]{Malik:2024nhy}%
  \BibitemOpen
  \bibfield  {author} {\bibinfo {author} {\bibfnamefont {Z.}~\bibnamefont {Malik}},\ }\href {\doibase 10.1016/j.aop.2025.170046} {\bibfield  {journal} {\bibinfo  {journal} {Annals Phys.}\ }\textbf {\bibinfo {volume} {479}},\ \bibinfo {pages} {170046} (\bibinfo {year} {2025}{\natexlab{c}})},\ \Eprint {http://arxiv.org/abs/2409.01561} {arXiv:2409.01561 [gr-qc]} \BibitemShut {NoStop}%
\bibitem [{\citenamefont {Dubinsky}(2025{\natexlab{d}})}]{Dubinsky:2025bvf}%
  \BibitemOpen
  \bibfield  {author} {\bibinfo {author} {\bibfnamefont {A.}~\bibnamefont {Dubinsky}},\ }\href {\doibase 10.1140/epjc/s10052-025-14671-3} {\bibfield  {journal} {\bibinfo  {journal} {Eur. Phys. J. C}\ }\textbf {\bibinfo {volume} {85}},\ \bibinfo {pages} {924} (\bibinfo {year} {2025}{\natexlab{d}})},\ \Eprint {http://arxiv.org/abs/2505.08545} {arXiv:2505.08545 [gr-qc]} \BibitemShut {NoStop}%
\bibitem [{\citenamefont {Skvortsova}(2024{\natexlab{d}})}]{Skvortsova:2024wly}%
  \BibitemOpen
  \bibfield  {author} {\bibinfo {author} {\bibfnamefont {M.}~\bibnamefont {Skvortsova}},\ }\href {\doibase 10.1134/S020228932470018X} {\bibfield  {journal} {\bibinfo  {journal} {Grav. Cosmol.}\ }\textbf {\bibinfo {volume} {30}},\ \bibinfo {pages} {279} (\bibinfo {year} {2024}{\natexlab{d}})},\ \Eprint {http://arxiv.org/abs/2405.15807} {arXiv:2405.15807 [gr-qc]} \BibitemShut {NoStop}%
\bibitem [{\citenamefont {Bolokhov}(2024{\natexlab{b}})}]{Bolokhov:2023bwm}%
  \BibitemOpen
  \bibfield  {author} {\bibinfo {author} {\bibfnamefont {S.~V.}\ \bibnamefont {Bolokhov}},\ }\href {\doibase 10.1103/PhysRevD.110.024010} {\bibfield  {journal} {\bibinfo  {journal} {Phys. Rev. D}\ }\textbf {\bibinfo {volume} {110}},\ \bibinfo {pages} {024010} (\bibinfo {year} {2024}{\natexlab{b}})},\ \Eprint {http://arxiv.org/abs/2311.05503} {arXiv:2311.05503 [gr-qc]} \BibitemShut {NoStop}%
\bibitem [{\citenamefont {Dubinsky}\ and\ \citenamefont {Zinhailo}(2025)}]{Dubinsky:2024nzo}%
  \BibitemOpen
  \bibfield  {author} {\bibinfo {author} {\bibfnamefont {A.}~\bibnamefont {Dubinsky}}\ and\ \bibinfo {author} {\bibfnamefont {A.~F.}\ \bibnamefont {Zinhailo}},\ }\href {\doibase 10.1209/0295-5075/adbc17} {\bibfield  {journal} {\bibinfo  {journal} {EPL}\ }\textbf {\bibinfo {volume} {149}},\ \bibinfo {pages} {69004} (\bibinfo {year} {2025})},\ \Eprint {http://arxiv.org/abs/2410.15232} {arXiv:2410.15232 [gr-qc]} \BibitemShut {NoStop}%
\bibitem [{\citenamefont {Gundlach}\ \emph {et~al.}(1994)\citenamefont {Gundlach}, \citenamefont {Price},\ and\ \citenamefont {Pullin}}]{Gundlach:1993tp}%
  \BibitemOpen
  \bibfield  {author} {\bibinfo {author} {\bibfnamefont {C.}~\bibnamefont {Gundlach}}, \bibinfo {author} {\bibfnamefont {R.~H.}\ \bibnamefont {Price}}, \ and\ \bibinfo {author} {\bibfnamefont {J.}~\bibnamefont {Pullin}},\ }\href {\doibase 10.1103/PhysRevD.49.883} {\bibfield  {journal} {\bibinfo  {journal} {Phys. Rev. D}\ }\textbf {\bibinfo {volume} {49}},\ \bibinfo {pages} {883} (\bibinfo {year} {1994})},\ \Eprint {http://arxiv.org/abs/gr-qc/9307009} {arXiv:gr-qc/9307009} \BibitemShut {NoStop}%
\bibitem [{\citenamefont {Ishihara}\ \emph {et~al.}(2008)\citenamefont {Ishihara}, \citenamefont {Kimura}, \citenamefont {Konoplya}, \citenamefont {Murata}, \citenamefont {Soda},\ and\ \citenamefont {Zhidenko}}]{Ishihara:2008re}%
  \BibitemOpen
  \bibfield  {author} {\bibinfo {author} {\bibfnamefont {H.}~\bibnamefont {Ishihara}}, \bibinfo {author} {\bibfnamefont {M.}~\bibnamefont {Kimura}}, \bibinfo {author} {\bibfnamefont {R.~A.}\ \bibnamefont {Konoplya}}, \bibinfo {author} {\bibfnamefont {K.}~\bibnamefont {Murata}}, \bibinfo {author} {\bibfnamefont {J.}~\bibnamefont {Soda}}, \ and\ \bibinfo {author} {\bibfnamefont {A.}~\bibnamefont {Zhidenko}},\ }\href {\doibase 10.1103/PhysRevD.77.084019} {\bibfield  {journal} {\bibinfo  {journal} {Phys. Rev. D}\ }\textbf {\bibinfo {volume} {77}},\ \bibinfo {pages} {084019} (\bibinfo {year} {2008})},\ \Eprint {http://arxiv.org/abs/0802.0655} {arXiv:0802.0655 [hep-th]} \BibitemShut {NoStop}%
\bibitem [{\citenamefont {Malik}(2025{\natexlab{d}})}]{Malik:2025ava}%
  \BibitemOpen
  \bibfield  {author} {\bibinfo {author} {\bibfnamefont {Z.}~\bibnamefont {Malik}},\ }\href {\doibase 10.1016/j.aop.2025.170238} {\bibfield  {journal} {\bibinfo  {journal} {Annals Phys.}\ }\textbf {\bibinfo {volume} {482}},\ \bibinfo {pages} {170238} (\bibinfo {year} {2025}{\natexlab{d}})},\ \Eprint {http://arxiv.org/abs/2504.12570} {arXiv:2504.12570 [gr-qc]} \BibitemShut {NoStop}%
\bibitem [{\citenamefont {L{\"u}tf{\"u}o{\u{g}}lu}(2025{\natexlab{c}})}]{Lutfuoglu:2025bsf}%
  \BibitemOpen
  \bibfield  {author} {\bibinfo {author} {\bibfnamefont {B.~C.}\ \bibnamefont {L{\"u}tf{\"u}o{\u{g}}lu}},\ }\href {\doibase 10.1016/j.physletb.2025.140026} {\bibfield  {journal} {\bibinfo  {journal} {Phys. Lett. B}\ }\textbf {\bibinfo {volume} {871}},\ \bibinfo {pages} {140026} (\bibinfo {year} {2025}{\natexlab{c}})},\ \Eprint {http://arxiv.org/abs/2508.13361} {arXiv:2508.13361 [gr-qc]} \BibitemShut {NoStop}%
\bibitem [{\citenamefont {Churilova}\ \emph {et~al.}(2020)\citenamefont {Churilova}, \citenamefont {Konoplya},\ and\ \citenamefont {Zhidenko}}]{Churilova:2019qph}%
  \BibitemOpen
  \bibfield  {author} {\bibinfo {author} {\bibfnamefont {M.~S.}\ \bibnamefont {Churilova}}, \bibinfo {author} {\bibfnamefont {R.~A.}\ \bibnamefont {Konoplya}}, \ and\ \bibinfo {author} {\bibfnamefont {A.}~\bibnamefont {Zhidenko}},\ }\href {\doibase 10.1016/j.physletb.2020.135207} {\bibfield  {journal} {\bibinfo  {journal} {Phys. Lett. B}\ }\textbf {\bibinfo {volume} {802}},\ \bibinfo {pages} {135207} (\bibinfo {year} {2020})},\ \Eprint {http://arxiv.org/abs/1911.05246} {arXiv:1911.05246 [gr-qc]} \BibitemShut {NoStop}%
\bibitem [{\citenamefont {Konoplya}\ and\ \citenamefont {Zhidenko}(2014)}]{Konoplya:2013sba}%
  \BibitemOpen
  \bibfield  {author} {\bibinfo {author} {\bibfnamefont {R.~A.}\ \bibnamefont {Konoplya}}\ and\ \bibinfo {author} {\bibfnamefont {A.}~\bibnamefont {Zhidenko}},\ }\href {\doibase 10.1103/PhysRevD.89.024011} {\bibfield  {journal} {\bibinfo  {journal} {Phys. Rev. D}\ }\textbf {\bibinfo {volume} {89}},\ \bibinfo {pages} {024011} (\bibinfo {year} {2014})},\ \Eprint {http://arxiv.org/abs/1309.7667} {arXiv:1309.7667 [hep-th]} \BibitemShut {NoStop}%
\bibitem [{\citenamefont {Qian}\ \emph {et~al.}(2022)\citenamefont {Qian}, \citenamefont {Lin}, \citenamefont {Shao}, \citenamefont {Wang},\ and\ \citenamefont {Yue}}]{Qian:2022kaq}%
  \BibitemOpen
  \bibfield  {author} {\bibinfo {author} {\bibfnamefont {W.-L.}\ \bibnamefont {Qian}}, \bibinfo {author} {\bibfnamefont {K.}~\bibnamefont {Lin}}, \bibinfo {author} {\bibfnamefont {C.-Y.}\ \bibnamefont {Shao}}, \bibinfo {author} {\bibfnamefont {B.}~\bibnamefont {Wang}}, \ and\ \bibinfo {author} {\bibfnamefont {R.-H.}\ \bibnamefont {Yue}},\ }\href {\doibase 10.1140/epjc/s10052-022-10910-z} {\bibfield  {journal} {\bibinfo  {journal} {Eur. Phys. J. C}\ }\textbf {\bibinfo {volume} {82}},\ \bibinfo {pages} {931} (\bibinfo {year} {2022})},\ \Eprint {http://arxiv.org/abs/2203.04477} {arXiv:2203.04477 [gr-qc]} \BibitemShut {NoStop}%
\bibitem [{\citenamefont {Varghese}\ and\ \citenamefont {Kuriakose}(2011)}]{Varghese:2011ku}%
  \BibitemOpen
  \bibfield  {author} {\bibinfo {author} {\bibfnamefont {N.}~\bibnamefont {Varghese}}\ and\ \bibinfo {author} {\bibfnamefont {V.~C.}\ \bibnamefont {Kuriakose}},\ }\href {\doibase 10.1007/s10714-011-1201-y} {\bibfield  {journal} {\bibinfo  {journal} {Gen. Rel. Grav.}\ }\textbf {\bibinfo {volume} {43}},\ \bibinfo {pages} {2757} (\bibinfo {year} {2011})},\ \Eprint {http://arxiv.org/abs/1011.6608} {arXiv:1011.6608 [gr-qc]} \BibitemShut {NoStop}%
\bibitem [{\citenamefont {Malik}(2024{\natexlab{b}})}]{Malik:2024bmp}%
  \BibitemOpen
  \bibfield  {author} {\bibinfo {author} {\bibfnamefont {Z.}~\bibnamefont {Malik}},\ }\href {\doibase 10.1515/zna-2024-0153} {\bibfield  {journal} {\bibinfo  {journal} {Z. Naturforsch. A}\ }\textbf {\bibinfo {volume} {79}},\ \bibinfo {pages} {1063} (\bibinfo {year} {2024}{\natexlab{b}})}\BibitemShut {NoStop}%
\bibitem [{\citenamefont {Konoplya}\ \emph {et~al.}(2020)\citenamefont {Konoplya}, \citenamefont {Zinhailo},\ and\ \citenamefont {Stuchlik}}]{Konoplya:2020jgt}%
  \BibitemOpen
  \bibfield  {author} {\bibinfo {author} {\bibfnamefont {R.~A.}\ \bibnamefont {Konoplya}}, \bibinfo {author} {\bibfnamefont {A.~F.}\ \bibnamefont {Zinhailo}}, \ and\ \bibinfo {author} {\bibfnamefont {Z.}~\bibnamefont {Stuchlik}},\ }\href {\doibase 10.1103/PhysRevD.102.044023} {\bibfield  {journal} {\bibinfo  {journal} {Phys. Rev. D}\ }\textbf {\bibinfo {volume} {102}},\ \bibinfo {pages} {044023} (\bibinfo {year} {2020})},\ \Eprint {http://arxiv.org/abs/2006.10462} {arXiv:2006.10462 [gr-qc]} \BibitemShut {NoStop}%
\bibitem [{\citenamefont {Dubinsky}(2024)}]{Dubinsky:2024jqi}%
  \BibitemOpen
  \bibfield  {author} {\bibinfo {author} {\bibfnamefont {A.}~\bibnamefont {Dubinsky}},\ }\href {\doibase 10.1209/0295-5075/ad51a3} {\bibfield  {journal} {\bibinfo  {journal} {EPL}\ }\textbf {\bibinfo {volume} {147}},\ \bibinfo {pages} {19003} (\bibinfo {year} {2024})},\ \Eprint {http://arxiv.org/abs/2403.01883} {arXiv:2403.01883 [gr-qc]} \BibitemShut {NoStop}%
\bibitem [{\citenamefont {Cuyubamba}\ \emph {et~al.}(2016)\citenamefont {Cuyubamba}, \citenamefont {Konoplya},\ and\ \citenamefont {Zhidenko}}]{Cuyubamba:2016cug}%
  \BibitemOpen
  \bibfield  {author} {\bibinfo {author} {\bibfnamefont {M.~A.}\ \bibnamefont {Cuyubamba}}, \bibinfo {author} {\bibfnamefont {R.~A.}\ \bibnamefont {Konoplya}}, \ and\ \bibinfo {author} {\bibfnamefont {A.}~\bibnamefont {Zhidenko}},\ }\href {\doibase 10.1103/PhysRevD.93.104053} {\bibfield  {journal} {\bibinfo  {journal} {Phys. Rev. D}\ }\textbf {\bibinfo {volume} {93}},\ \bibinfo {pages} {104053} (\bibinfo {year} {2016})},\ \Eprint {http://arxiv.org/abs/1604.03604} {arXiv:1604.03604 [gr-qc]} \BibitemShut {NoStop}%
\bibitem [{\citenamefont {Bolokhov}(2024{\natexlab{c}})}]{Bolokhov:2023ruj}%
  \BibitemOpen
  \bibfield  {author} {\bibinfo {author} {\bibfnamefont {S.~V.}\ \bibnamefont {Bolokhov}},\ }\href {\doibase 10.1103/PhysRevD.109.064017} {\bibfield  {journal} {\bibinfo  {journal} {Phys. Rev. D}\ }\textbf {\bibinfo {volume} {109}},\ \bibinfo {pages} {064017} (\bibinfo {year} {2024}{\natexlab{c}})}\BibitemShut {NoStop}%
\bibitem [{\citenamefont {L{\"u}tf{\"u}o{\u{g}}lu}(2025{\natexlab{d}})}]{Lutfuoglu:2025hwh}%
  \BibitemOpen
  \bibfield  {author} {\bibinfo {author} {\bibfnamefont {B.~C.}\ \bibnamefont {L{\"u}tf{\"u}o{\u{g}}lu}},\ }\href {\doibase 10.1088/1475-7516/2025/06/057} {\bibfield  {journal} {\bibinfo  {journal} {JCAP}\ }\textbf {\bibinfo {volume} {06}},\ \bibinfo {pages} {057} (\bibinfo {year} {2025}{\natexlab{d}})},\ \Eprint {http://arxiv.org/abs/2504.09323} {arXiv:2504.09323 [gr-qc]} \BibitemShut {NoStop}%
\bibitem [{\citenamefont {L{\"u}tf{\"u}o{\u{g}}lu}(2025{\natexlab{e}})}]{Lutfuoglu:2025qkt}%
  \BibitemOpen
  \bibfield  {author} {\bibinfo {author} {\bibfnamefont {B.~C.}\ \bibnamefont {L{\"u}tf{\"u}o{\u{g}}lu}},\ }\href {\doibase 10.1140/epjc/s10052-025-14839-x} {\bibfield  {journal} {\bibinfo  {journal} {Eur. Phys. J. C}\ }\textbf {\bibinfo {volume} {85}},\ \bibinfo {pages} {1076} (\bibinfo {year} {2025}{\natexlab{e}})},\ \Eprint {http://arxiv.org/abs/2508.19194} {arXiv:2508.19194 [gr-qc]} \BibitemShut {NoStop}%
\bibitem [{\citenamefont {Konoplya}\ and\ \citenamefont {Zhidenko}(2024{\natexlab{b}})}]{Konoplya:2024lir}%
  \BibitemOpen
  \bibfield  {author} {\bibinfo {author} {\bibfnamefont {R.~A.}\ \bibnamefont {Konoplya}}\ and\ \bibinfo {author} {\bibfnamefont {A.}~\bibnamefont {Zhidenko}},\ }\href {\doibase 10.1088/1475-7516/2024/09/068} {\bibfield  {journal} {\bibinfo  {journal} {JCAP}\ }\textbf {\bibinfo {volume} {09}},\ \bibinfo {pages} {068} (\bibinfo {year} {2024}{\natexlab{b}})},\ \Eprint {http://arxiv.org/abs/2406.11694} {arXiv:2406.11694 [gr-qc]} \BibitemShut {NoStop}%
\bibitem [{\citenamefont {L{\"u}tf{\"u}o{\u{g}}lu}(2025{\natexlab{f}})}]{Lutfuoglu:2025ldc}%
  \BibitemOpen
  \bibfield  {author} {\bibinfo {author} {\bibfnamefont {B.~C.}\ \bibnamefont {L{\"u}tf{\"u}o{\u{g}}lu}},\ }\href {\doibase 10.53941/ijgtp.2025.100004} {\bibfield  {journal} {\bibinfo  {journal} {Int. J. Grav. Theor. Phys}\ }\textbf {\bibinfo {volume} {1}},\ \bibinfo {pages} {4} (\bibinfo {year} {2025}{\natexlab{f}})},\ \Eprint {http://arxiv.org/abs/2507.09246} {arXiv:2507.09246 [gr-qc]} \BibitemShut {NoStop}%
\bibitem [{\citenamefont {Malik}(2025{\natexlab{e}})}]{Malik:2024cgb}%
  \BibitemOpen
  \bibfield  {author} {\bibinfo {author} {\bibfnamefont {Z.}~\bibnamefont {Malik}},\ }\href {\doibase 10.1088/1475-7516/2025/04/042} {\bibfield  {journal} {\bibinfo  {journal} {JCAP}\ }\textbf {\bibinfo {volume} {04}},\ \bibinfo {pages} {042} (\bibinfo {year} {2025}{\natexlab{e}})},\ \Eprint {http://arxiv.org/abs/2412.19443} {arXiv:2412.19443 [gr-qc]} \BibitemShut {NoStop}%
\bibitem [{\citenamefont {Bolokhov}\ and\ \citenamefont {Skvortsova}(2025{\natexlab{d}})}]{Bolokhov:2024otn}%
  \BibitemOpen
  \bibfield  {author} {\bibinfo {author} {\bibfnamefont {S.~V.}\ \bibnamefont {Bolokhov}}\ and\ \bibinfo {author} {\bibfnamefont {M.}~\bibnamefont {Skvortsova}},\ }\href {\doibase 10.1088/1475-7516/2025/04/025} {\bibfield  {journal} {\bibinfo  {journal} {JCAP}\ }\textbf {\bibinfo {volume} {04}},\ \bibinfo {pages} {025} (\bibinfo {year} {2025}{\natexlab{d}})},\ \Eprint {http://arxiv.org/abs/2412.11166} {arXiv:2412.11166 [gr-qc]} \BibitemShut {NoStop}%
\bibitem [{\citenamefont {Han}\ and\ \citenamefont {Gwak}(2025)}]{Han:2025cal}%
  \BibitemOpen
  \bibfield  {author} {\bibinfo {author} {\bibfnamefont {H.}~\bibnamefont {Han}}\ and\ \bibinfo {author} {\bibfnamefont {B.}~\bibnamefont {Gwak}},\ }\href@noop {} {\  (\bibinfo {year} {2025})},\ \Eprint {http://arxiv.org/abs/2508.12989} {arXiv:2508.12989 [gr-qc]} \BibitemShut {NoStop}%
\bibitem [{\citenamefont {Abbott}\ \emph {et~al.}(2025)\citenamefont {Abbott} \emph {et~al.}}]{LIGOScientific:2021sio}%
  \BibitemOpen
  \bibfield  {author} {\bibinfo {author} {\bibfnamefont {R.}~\bibnamefont {Abbott}} \emph {et~al.} (\bibinfo {collaboration} {LIGO Scientific, VIRGO, KAGRA}),\ }\href {\doibase 10.1103/PhysRevD.112.084080} {\bibfield  {journal} {\bibinfo  {journal} {Phys. Rev. D}\ }\textbf {\bibinfo {volume} {112}},\ \bibinfo {pages} {084080} (\bibinfo {year} {2025})},\ \Eprint {http://arxiv.org/abs/2112.06861} {arXiv:2112.06861 [gr-qc]} \BibitemShut {NoStop}%
\bibitem [{\citenamefont {Ghosh}\ \emph {et~al.}(2021)\citenamefont {Ghosh}, \citenamefont {Brito},\ and\ \citenamefont {Buonanno}}]{Ghosh:2021mrv}%
  \BibitemOpen
  \bibfield  {author} {\bibinfo {author} {\bibfnamefont {A.}~\bibnamefont {Ghosh}}, \bibinfo {author} {\bibfnamefont {R.}~\bibnamefont {Brito}}, \ and\ \bibinfo {author} {\bibfnamefont {A.}~\bibnamefont {Buonanno}},\ }\href {\doibase 10.1103/PhysRevD.103.124041} {\bibfield  {journal} {\bibinfo  {journal} {Phys. Rev. D}\ }\textbf {\bibinfo {volume} {103}},\ \bibinfo {pages} {124041} (\bibinfo {year} {2021})},\ \Eprint {http://arxiv.org/abs/2104.01906} {arXiv:2104.01906 [gr-qc]} \BibitemShut {NoStop}%
\bibitem [{\citenamefont {Cotesta}\ \emph {et~al.}(2022)\citenamefont {Cotesta}, \citenamefont {Carullo}, \citenamefont {Berti},\ and\ \citenamefont {Cardoso}}]{Cotesta:2022pci}%
  \BibitemOpen
  \bibfield  {author} {\bibinfo {author} {\bibfnamefont {R.}~\bibnamefont {Cotesta}}, \bibinfo {author} {\bibfnamefont {G.}~\bibnamefont {Carullo}}, \bibinfo {author} {\bibfnamefont {E.}~\bibnamefont {Berti}}, \ and\ \bibinfo {author} {\bibfnamefont {V.}~\bibnamefont {Cardoso}},\ }\href {\doibase 10.1103/PhysRevLett.129.111102} {\bibfield  {journal} {\bibinfo  {journal} {Phys. Rev. Lett.}\ }\textbf {\bibinfo {volume} {129}},\ \bibinfo {pages} {111102} (\bibinfo {year} {2022})},\ \Eprint {http://arxiv.org/abs/2201.00822} {arXiv:2201.00822 [gr-qc]} \BibitemShut {NoStop}%
\bibitem [{\citenamefont {Bhagwat}\ \emph {et~al.}(2023)\citenamefont {Bhagwat}, \citenamefont {Pacilio}, \citenamefont {Pani},\ and\ \citenamefont {Mapelli}}]{Bhagwat:2023jwv}%
  \BibitemOpen
  \bibfield  {author} {\bibinfo {author} {\bibfnamefont {S.}~\bibnamefont {Bhagwat}}, \bibinfo {author} {\bibfnamefont {C.}~\bibnamefont {Pacilio}}, \bibinfo {author} {\bibfnamefont {P.}~\bibnamefont {Pani}}, \ and\ \bibinfo {author} {\bibfnamefont {M.}~\bibnamefont {Mapelli}},\ }\href {\doibase 10.1103/PhysRevD.108.043019} {\bibfield  {journal} {\bibinfo  {journal} {Phys. Rev. D}\ }\textbf {\bibinfo {volume} {108}},\ \bibinfo {pages} {043019} (\bibinfo {year} {2023})},\ \Eprint {http://arxiv.org/abs/2304.02283} {arXiv:2304.02283 [gr-qc]} \BibitemShut {NoStop}%
\bibitem [{\citenamefont {Berti}\ \emph {et~al.}(2025)\citenamefont {Berti} \emph {et~al.}}]{Berti:2025hly}%
  \BibitemOpen
  \bibfield  {author} {\bibinfo {author} {\bibfnamefont {E.}~\bibnamefont {Berti}} \emph {et~al.},\ }\href@noop {} {\  (\bibinfo {year} {2025})},\ \Eprint {http://arxiv.org/abs/2505.23895} {arXiv:2505.23895 [gr-qc]} \BibitemShut {NoStop}%
\bibitem [{\citenamefont {Abac}\ \emph {et~al.}(2025)\citenamefont {Abac} \emph {et~al.}}]{KAGRA:2025oiz}%
  \BibitemOpen
  \bibfield  {author} {\bibinfo {author} {\bibfnamefont {A.~G.}\ \bibnamefont {Abac}} \emph {et~al.} (\bibinfo {collaboration} {KAGRA, Virgo, LIGO Scientific}),\ }\href {\doibase 10.1103/kw5g-d732} {\bibfield  {journal} {\bibinfo  {journal} {Phys. Rev. Lett.}\ }\textbf {\bibinfo {volume} {135}},\ \bibinfo {pages} {111403} (\bibinfo {year} {2025})},\ \Eprint {http://arxiv.org/abs/2509.08054} {arXiv:2509.08054 [gr-qc]} \BibitemShut {NoStop}%
\bibitem [{\citenamefont {Cardoso}\ \emph {et~al.}(2009)\citenamefont {Cardoso}, \citenamefont {Miranda}, \citenamefont {Berti}, \citenamefont {Witek},\ and\ \citenamefont {Zanchin}}]{Cardoso:2008bp}%
  \BibitemOpen
  \bibfield  {author} {\bibinfo {author} {\bibfnamefont {V.}~\bibnamefont {Cardoso}}, \bibinfo {author} {\bibfnamefont {A.~S.}\ \bibnamefont {Miranda}}, \bibinfo {author} {\bibfnamefont {E.}~\bibnamefont {Berti}}, \bibinfo {author} {\bibfnamefont {H.}~\bibnamefont {Witek}}, \ and\ \bibinfo {author} {\bibfnamefont {V.~T.}\ \bibnamefont {Zanchin}},\ }\href {\doibase 10.1103/PhysRevD.79.064016} {\bibfield  {journal} {\bibinfo  {journal} {Phys. Rev. D}\ }\textbf {\bibinfo {volume} {79}},\ \bibinfo {pages} {064016} (\bibinfo {year} {2009})},\ \Eprint {http://arxiv.org/abs/0812.1806} {arXiv:0812.1806 [hep-th]} \BibitemShut {NoStop}%
\bibitem [{\citenamefont {Khanna}\ and\ \citenamefont {Price}(2017)}]{Khanna:2016yow}%
  \BibitemOpen
  \bibfield  {author} {\bibinfo {author} {\bibfnamefont {G.}~\bibnamefont {Khanna}}\ and\ \bibinfo {author} {\bibfnamefont {R.~H.}\ \bibnamefont {Price}},\ }\href {\doibase 10.1103/PhysRevD.95.081501} {\bibfield  {journal} {\bibinfo  {journal} {Phys. Rev. D}\ }\textbf {\bibinfo {volume} {95}},\ \bibinfo {pages} {081501} (\bibinfo {year} {2017})},\ \Eprint {http://arxiv.org/abs/1609.00083} {arXiv:1609.00083 [gr-qc]} \BibitemShut {NoStop}%
\bibitem [{\citenamefont {Konoplya}\ and\ \citenamefont {Stuchl{\'\i}k}(2017)}]{Konoplya:2017wot}%
  \BibitemOpen
  \bibfield  {author} {\bibinfo {author} {\bibfnamefont {R.~A.}\ \bibnamefont {Konoplya}}\ and\ \bibinfo {author} {\bibfnamefont {Z.}~\bibnamefont {Stuchl{\'\i}k}},\ }\href {\doibase 10.1016/j.physletb.2017.06.015} {\bibfield  {journal} {\bibinfo  {journal} {Phys. Lett. B}\ }\textbf {\bibinfo {volume} {771}},\ \bibinfo {pages} {597} (\bibinfo {year} {2017})},\ \Eprint {http://arxiv.org/abs/1705.05928} {arXiv:1705.05928 [gr-qc]} \BibitemShut {NoStop}%
\bibitem [{\citenamefont {Bolokhov}(2024{\natexlab{d}})}]{Bolokhov:2023dxq}%
  \BibitemOpen
  \bibfield  {author} {\bibinfo {author} {\bibfnamefont {S.~V.}\ \bibnamefont {Bolokhov}},\ }\href {\doibase 10.1016/j.physletb.2024.138879} {\bibfield  {journal} {\bibinfo  {journal} {Phys. Lett. B}\ }\textbf {\bibinfo {volume} {856}},\ \bibinfo {pages} {138879} (\bibinfo {year} {2024}{\natexlab{d}})},\ \Eprint {http://arxiv.org/abs/2310.12326} {arXiv:2310.12326 [gr-qc]} \BibitemShut {NoStop}%
\bibitem [{\citenamefont {Konoplya}\ and\ \citenamefont {Zhidenko}(2017)}]{Konoplya:2017lhs}%
  \BibitemOpen
  \bibfield  {author} {\bibinfo {author} {\bibfnamefont {R.~A.}\ \bibnamefont {Konoplya}}\ and\ \bibinfo {author} {\bibfnamefont {A.}~\bibnamefont {Zhidenko}},\ }\href {\doibase 10.1088/1475-7516/2017/05/050} {\bibfield  {journal} {\bibinfo  {journal} {JCAP}\ }\textbf {\bibinfo {volume} {05}},\ \bibinfo {pages} {050} (\bibinfo {year} {2017})},\ \Eprint {http://arxiv.org/abs/1705.01656} {arXiv:1705.01656 [hep-th]} \BibitemShut {NoStop}%
\bibitem [{\citenamefont {Konoplya}\ \emph {et~al.}(2025{\natexlab{a}})\citenamefont {Konoplya}, \citenamefont {Spina},\ and\ \citenamefont {Zhidenko}}]{Konoplya:2025afm}%
  \BibitemOpen
  \bibfield  {author} {\bibinfo {author} {\bibfnamefont {R.~A.}\ \bibnamefont {Konoplya}}, \bibinfo {author} {\bibfnamefont {A.}~\bibnamefont {Spina}}, \ and\ \bibinfo {author} {\bibfnamefont {A.}~\bibnamefont {Zhidenko}},\ }\href {\doibase 10.1103/xhtc-9cf4} {\bibfield  {journal} {\bibinfo  {journal} {Phys. Rev. D}\ }\textbf {\bibinfo {volume} {112}},\ \bibinfo {pages} {024060} (\bibinfo {year} {2025}{\natexlab{a}})},\ \Eprint {http://arxiv.org/abs/2505.01128} {arXiv:2505.01128 [gr-qc]} \BibitemShut {NoStop}%
\bibitem [{\citenamefont {Konoplya}\ \emph {et~al.}(2025{\natexlab{b}})\citenamefont {Konoplya}, \citenamefont {Khrabustovskyi}, \citenamefont {K{\v{r}}{\'\i}{\v{z}}},\ and\ \citenamefont {Zhidenko}}]{Konoplya:2025mvj}%
  \BibitemOpen
  \bibfield  {author} {\bibinfo {author} {\bibfnamefont {R.~A.}\ \bibnamefont {Konoplya}}, \bibinfo {author} {\bibfnamefont {A.}~\bibnamefont {Khrabustovskyi}}, \bibinfo {author} {\bibfnamefont {J.}~\bibnamefont {K{\v{r}}{\'\i}{\v{z}}}}, \ and\ \bibinfo {author} {\bibfnamefont {A.}~\bibnamefont {Zhidenko}},\ }\href {\doibase 10.1088/1475-7516/2025/04/062} {\bibfield  {journal} {\bibinfo  {journal} {JCAP}\ }\textbf {\bibinfo {volume} {04}},\ \bibinfo {pages} {062} (\bibinfo {year} {2025}{\natexlab{b}})},\ \Eprint {http://arxiv.org/abs/2501.16134} {arXiv:2501.16134 [gr-qc]} \BibitemShut {NoStop}%
\bibitem [{\citenamefont {Konoplya}(2023)}]{Konoplya:2022gjp}%
  \BibitemOpen
  \bibfield  {author} {\bibinfo {author} {\bibfnamefont {R.~A.}\ \bibnamefont {Konoplya}},\ }\href {\doibase 10.1016/j.physletb.2023.137674} {\bibfield  {journal} {\bibinfo  {journal} {Phys. Lett. B}\ }\textbf {\bibinfo {volume} {838}},\ \bibinfo {pages} {137674} (\bibinfo {year} {2023})},\ \Eprint {http://arxiv.org/abs/2210.08373} {arXiv:2210.08373 [gr-qc]} \BibitemShut {NoStop}%
\bibitem [{\citenamefont {Konoplya}\ and\ \citenamefont {Zhidenko}(2024{\natexlab{c}})}]{Konoplya:2022pbc}%
  \BibitemOpen
  \bibfield  {author} {\bibinfo {author} {\bibfnamefont {R.~A.}\ \bibnamefont {Konoplya}}\ and\ \bibinfo {author} {\bibfnamefont {A.}~\bibnamefont {Zhidenko}},\ }\href {\doibase 10.1016/j.jheap.2024.10.015} {\bibfield  {journal} {\bibinfo  {journal} {JHEAp}\ }\textbf {\bibinfo {volume} {44}},\ \bibinfo {pages} {419} (\bibinfo {year} {2024}{\natexlab{c}})},\ \Eprint {http://arxiv.org/abs/2209.00679} {arXiv:2209.00679 [gr-qc]} \BibitemShut {NoStop}%
\bibitem [{\citenamefont {Konoplya}\ \emph {et~al.}(2022)\citenamefont {Konoplya}, \citenamefont {Zinhailo}, \citenamefont {Kunz}, \citenamefont {Stuchlik},\ and\ \citenamefont {Zhidenko}}]{Konoplya:2022hll}%
  \BibitemOpen
  \bibfield  {author} {\bibinfo {author} {\bibfnamefont {R.~A.}\ \bibnamefont {Konoplya}}, \bibinfo {author} {\bibfnamefont {A.~F.}\ \bibnamefont {Zinhailo}}, \bibinfo {author} {\bibfnamefont {J.}~\bibnamefont {Kunz}}, \bibinfo {author} {\bibfnamefont {Z.}~\bibnamefont {Stuchlik}}, \ and\ \bibinfo {author} {\bibfnamefont {A.}~\bibnamefont {Zhidenko}},\ }\href {\doibase 10.1088/1475-7516/2022/10/091} {\bibfield  {journal} {\bibinfo  {journal} {JCAP}\ }\textbf {\bibinfo {volume} {10}},\ \bibinfo {pages} {091} (\bibinfo {year} {2022})},\ \Eprint {http://arxiv.org/abs/2206.14714} {arXiv:2206.14714 [gr-qc]} \BibitemShut {NoStop}%
\bibitem [{\citenamefont {Leaver}(1985)}]{Leaver:1985ax}%
  \BibitemOpen
  \bibfield  {author} {\bibinfo {author} {\bibfnamefont {E.~W.}\ \bibnamefont {Leaver}},\ }\href {\doibase 10.1098/rspa.1985.0119} {\bibfield  {journal} {\bibinfo  {journal} {Proc. Roy. Soc. Lond. A}\ }\textbf {\bibinfo {volume} {402}},\ \bibinfo {pages} {285} (\bibinfo {year} {1985})}\BibitemShut {NoStop}%
\end{thebibliography}%

\end{document}